\documentclass[reprint,amsmath,amssymb, aps,longbibliography]{revtex4-2}

\usepackage{graphicx}
\usepackage{dcolumn}
\usepackage{bm}
\usepackage[caption=false]{subfig} 
\usepackage{physics}
\usepackage{tikz}
\usepackage{siunitx}
\usepackage{placeins}

\newcommand{\kth}{\mathbf{K}_\theta}
\newcommand{\meV}{\,\si{meV}}

\usepackage[colorlinks,linkcolor=blue,citecolor=blue,urlcolor=blue,filecolor=black]{hyperref}
\usepackage[capitalise,noabbrev,nameinlink]{cleveref}
\crefname{appendix}{Appendix}{Appendices}
\Crefname{appendix}{Appendix}{Appendices}
\crefname{equation}{Eq.}{Eqs.}
\Crefname{equation}{Equation}{Equations}
\crefname{figure}{Fig.}{Figs.}
\Crefname{figure}{Figure}{Figures}

\begin{document}
\title{Momentum-resolved spectroscopy of superconductivity with the quantum twisting microscope}

\author{Yuval Waschitz}
\email{yuval.waschitz@weizmann.ac.il}
\author{Ady Stern}
\author{Yuval Oreg}

\affiliation{Department of Condensed Matter Physics, Weizmann Institute of Science, Rehovot, Israel 76100}

\date{April 15, 2026}

\begin{abstract}
We develop a theoretical framework for probing superconductivity with momentum resolution using the quantum twisting microscope (QTM), a planar tunneling device where a graphene tip is rotated relative to a two-dimensional sample. Because of in-plane momentum conservation, the QTM directly measures the superconducting spectral function along well-defined trajectories in momentum space. The relative intensities of electron and hole excitations encode the Bogoliubov coherence factors, revealing the momentum dependence of the pairing magnitude. Three $C_{3z}$-related tunneling channels enable direct detection of rotational symmetry breaking, as well as nodal points in the superconducting order parameter. We apply our framework to superconductivity within the Bistritzer-MacDonald model of noninteracting electrons and the topological heavy-fermion model, which accounts for electron-electron interactions. Together, these capabilities establish the QTM as a direct probe of the pairing symmetry and microscopic origin of superconductivity in two-dimensional materials.
\end{abstract}

\maketitle

The superconducting pairing potential encodes the microscopic mechanism responsible for electron pairing. In conventional superconductors, the pairing is described by the BCS theory and is assumed to be momentum independent~\cite{bardeen1957theory}. In contrast, in the high-$T_c$ cuprates and in many unconventional superconductors the pairing potential varies with momentum and may exhibit nodes~\cite{tsuei2000pairing,sigrist1991phenomenological,sigrist2005introduction}. Recent discoveries of superconductivity in graphene systems, including magic-angle twisted bilayer graphene (MATBG)~\cite{cao2018unconventional,yankowitz2019tuning,lu2019superconductors,park2021tunable,hao2021electric,zhou2021superconductivity,zhou2022isospin}, have revealed signatures of nematic and nodal pairing~\cite{cao2021nematicity,zhang2025angular,banerjee2025superfluid,tanaka2025superfluid,oh2021evidence,kim2022evidence,park2025simultaneous}, suggesting physics beyond conventional BCS theory~\cite{lake2022pairing} and prompting extensive theoretical proposals for pairing symmetries and mechanisms~\cite{kennes2018strong,peltonen2018mean,isobe2018unconventional,liu2018chiral,wu2018theory,wu2019topological,lian2019twisted,huang2019antiferromagnetically,kozii2019nematic,cea2021coulomb,wang2021topological,shavit2021theory,khalaf2021charged,yu2022euler,wang2024molecular,liu2025nodal}.

\begin{figure}[t]
  \centering
  \subfloat{%
    \begin{tikzpicture}
      \node[inner sep=0] (img) {\includegraphics[width=0.6\columnwidth]{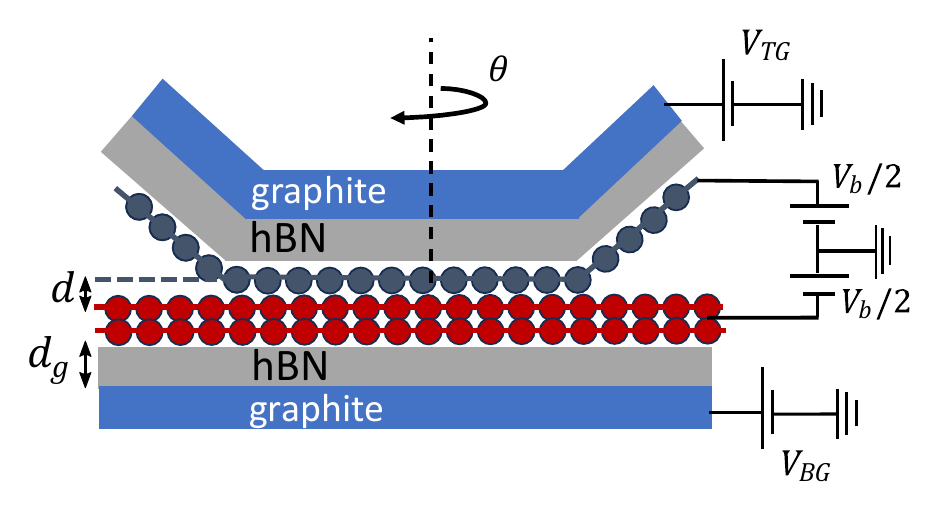}};
      \node[anchor=north west, xshift=0pt, yshift=2pt]  at (img.north west) {\textbf{(a)}};
    \end{tikzpicture}%
    \label{fig:QTM_a}%
  }%
  \subfloat{%
    \begin{tikzpicture}
      \node[inner sep=0] (img) {\includegraphics[width=0.4\columnwidth]{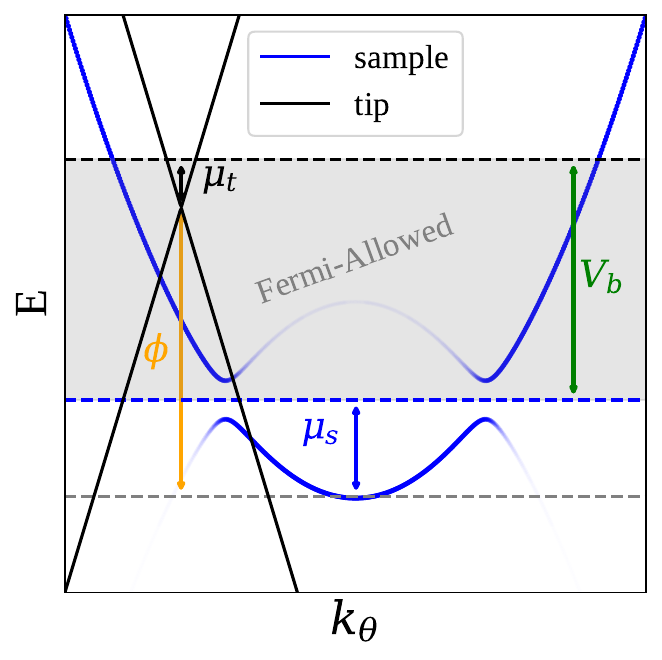}};
      \node[anchor=north west, xshift=6pt, yshift=12pt]  at (img.north west) {\textbf{(b)}};
    \end{tikzpicture}%
    \label{fig:QTM_b}%
  }%
\caption{(a) Schematic diagram of the QTM junction (adapted from~\cite{theoryphonons}), consisting of tip (blue circles) and sample (red circles) layers with independently tunable gate and bias voltages. The top layer can be rotated relative to the bottom in a controlled manner.
(b) Band alignment of a monolayer graphene tip with a parabolic sample band. A superconducting gap is introduced in the sample; the color intensity reflects the electron spectral weight. The chemical potentials of tip and sample ($\mu_T,\mu_S$), the electrostatic shift $\phi$, applied bias $V_b$, and the corresponding tunneling window (gray rectangle) are indicated.}
\label{fig:QTM_subfigs}

\end{figure}
In this work, we show that the recently developed quantum twisting microscope (QTM) \cite{inbar2023quantum} provides direct access to the magnitude of the superconducting pairing potential $|\Delta_{\bf k}|$ in momentum space. In particular, it can detect spontaneous breaking of  the $C_{3z}$ symmetry of a hexagonal lattice, resolve coherence factors with momentum resolution, and reveal nodal excitations and their momenta through zero-bias conductance.

The QTM~\cite{inbar2023quantum} is a tunneling device in which a two-dimensional crystalline tip is rotated relative to a two-dimensional sample, forming a planar junction (see~\cref{fig:QTM_a}).
Because the junction has a large contact area and the tip retains its crystalline order, tunneling processes conserve in-plane momentum. 
This makes the tunneling spectrum directly sensitive to the sample band structure and single-particle spectral function in energy--momentum space. Separate tip and sample gate voltages, together with a bias voltage ($V_b$), control the chemical potentials of the tip and sample ($\mu_T$, $\mu_S$), and the electrostatic shift between them ($\phi$) (see~\cref{fig:QTM_a,fig:QTM_b}).

Theoretically, the QTM has been proposed as a probe of diverse correlated and collective phenomena~\cite{qtmspins,qtmmagnet,wei2025theory,theoryphonons,diracscan,xiao2023probing}, while experiments have demonstrated its ability to map band structures~\cite{inbar2023quantum,lee2025revealing}, phonons~\cite{qtmphonons}, and interaction effects in MATBG~\cite{xiao2025interacting}, establishing it as a powerful and promising tool for momentum-resolved spectroscopy.

To resolve the sample spectral function, momentum conservation alone is not sufficient; localization is required in both momentum and energy, such that the tunneling current involves only a narrow range of momenta and energy. 
This is provided by the sharp linear dispersion of a monolayer-graphene (MLG) tip with circular Fermi surface set by~$\mu_T$. 
We treat $\mu_T$ and $\mu_S$ as bias-independent, so $V_b$ only shifts the electrostatic potential~$\phi$, practically, we assume this can be achieved with appropriate gate compensation (see discussion around \cref{eq:constphi} in \cite{SM}). \nocite{slizovskiy2021out, boschi2024built,mahan1990many, song_allmagic,cualuguaru2023twisted,dynestwo, dynesexp1, dynesexp2, schafferpairing}
As $\phi$ varies, the tip Dirac cone sweeps through the sample bands, yielding sharp differential conductance features at their crossing that trace the sample spectral function in the mBZ~\cite{inbar2023quantum,diracscan,SM}.

We evaluate the differential conductance ($I''\equiv d^2I/dV_b^2$) analytically, in the case where $\mu_T=0$ and the sample is superconducting. In the limit of infinite quasiparticle lifetime, zero temperature and sample bands which are flat relative to those of the tip, we obtain~\cite{SM}
\begin{multline}
I''(\theta,V_b)\approx
  \frac{\Omega e ^3 }{\hbar^{3} v_D^2}|T(\kth)|^2 \times
\\\left[|v_{\kth}|^2 \,\delta(eV_b -E_{\kth}) -| u_{\kth}|^2 \,\delta(eV_b + E_{\kth})\right],
\label{eq:tun_sc}
\end{multline}
where $\Omega$ is the junction area, $v_D$ is the Dirac group velocity in the graphene tip; and $T(\mathbf{K}_\theta)$ is the tunneling matrix element between states near the tip Dirac point. The momentum $\mathbf{K}_\theta$ at which the sample is being probed is set by the rotation angle $\theta$ of the tip. More detailed definitions are given in the discussion leading to \cref{eq:tamp_kth} in Ref.~\cite{SM}. In \cref{eq:tun_sc} the energy $E_{\mathbf{k}}$ is the Bogoliubov quasiparticle energy and $u_{\mathbf{k}},v_{\mathbf{k}}$ are the coherence factors. 
For a time-reversal-symmetric (TRS) normal state with dispersion $\xi_{\mathbf{k}}$, 
$E_{\mathbf{k}}=\sqrt{\xi_{\mathbf{k}}^{2}+|\Delta_{\mathbf{k}}|^{2}}$, assuming singlet or unitary triplet pairing, 
so $|\Delta_{\mathbf{k}}|^2$ is spin independent and treated as a scalar~\cite{mineev1999introduction,sigrist2005introduction}. \cref{eq:tun_sc} implies that the Bogoliubov excitation spectrum can be mapped along the trajectory $\mathbf{K}_\theta$ by varying $V_b$ and $\theta$.
Setting $\mu_T = 0$ is not essential and is chosen for convenience, as it allows measurement of both electron and hole excitations without changing the sign of $\mu_T$~\cite{SM}. A similar expression for the first current derivative $I'(\theta,V_b)$ can be obtained by fixing $\phi$ and varying $\mu_T$ for small tip doping (see~\cref{eq:dIdV_mut} in Ref.~\cite{SM}).

\begin{figure}[t]
  \centering
    \begin{tikzpicture}
      \node[inner sep=0] (img) {\includegraphics[width=0.95\columnwidth]{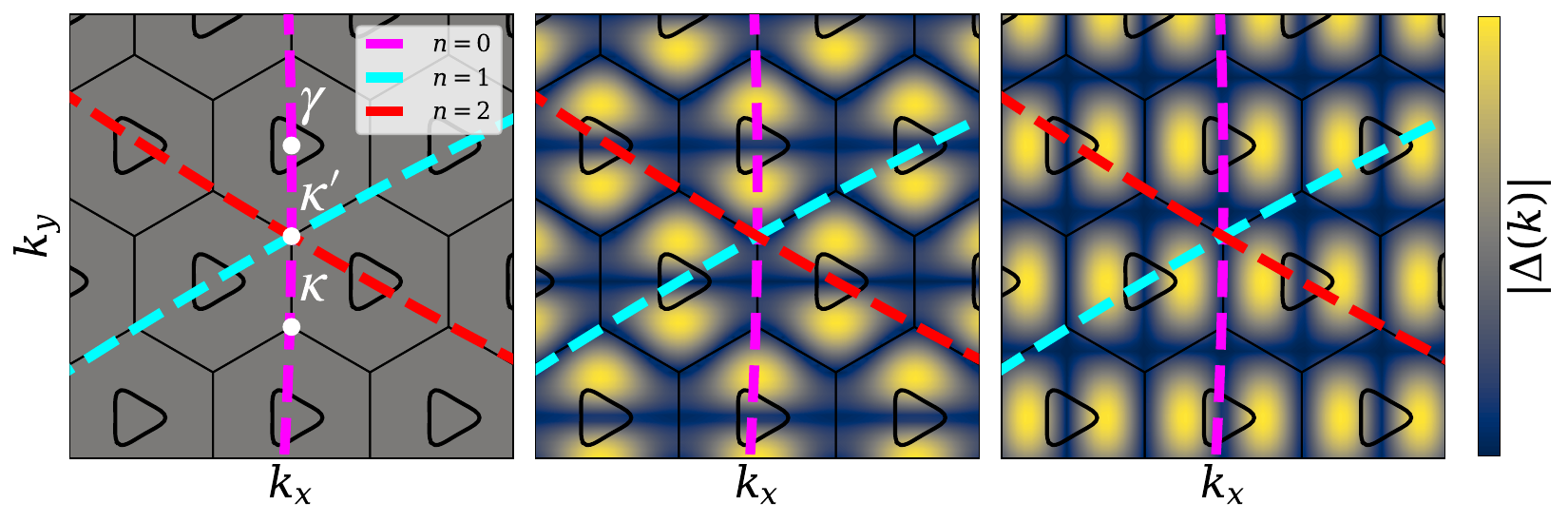}};
      \node[anchor=north west, xshift=8pt, yshift=12pt] 
        at (img.north west) {\textbf{(a)}}; 
      \node[anchor=north west, xshift=\columnwidth/3-4pt, yshift=12pt] 
        at (img.north west) {\textbf{(b)}};
      \node[anchor=north west, xshift=\columnwidth*2/3-18pt, yshift=12pt] 
        at (img.north west) {\textbf{(c)}}; 
    \end{tikzpicture}%
  \subfloat{\label{fig:mbzs}}\hspace{0pt}%
  \subfloat{\label{fig:mbzpx}}\hspace{0pt}%
  \subfloat{\label{fig:mbzpy}}\hspace{0pt}%
   \vspace{-1em}
    \begin{tikzpicture}
      \node[inner sep=0] (img) {\includegraphics[width=\columnwidth]{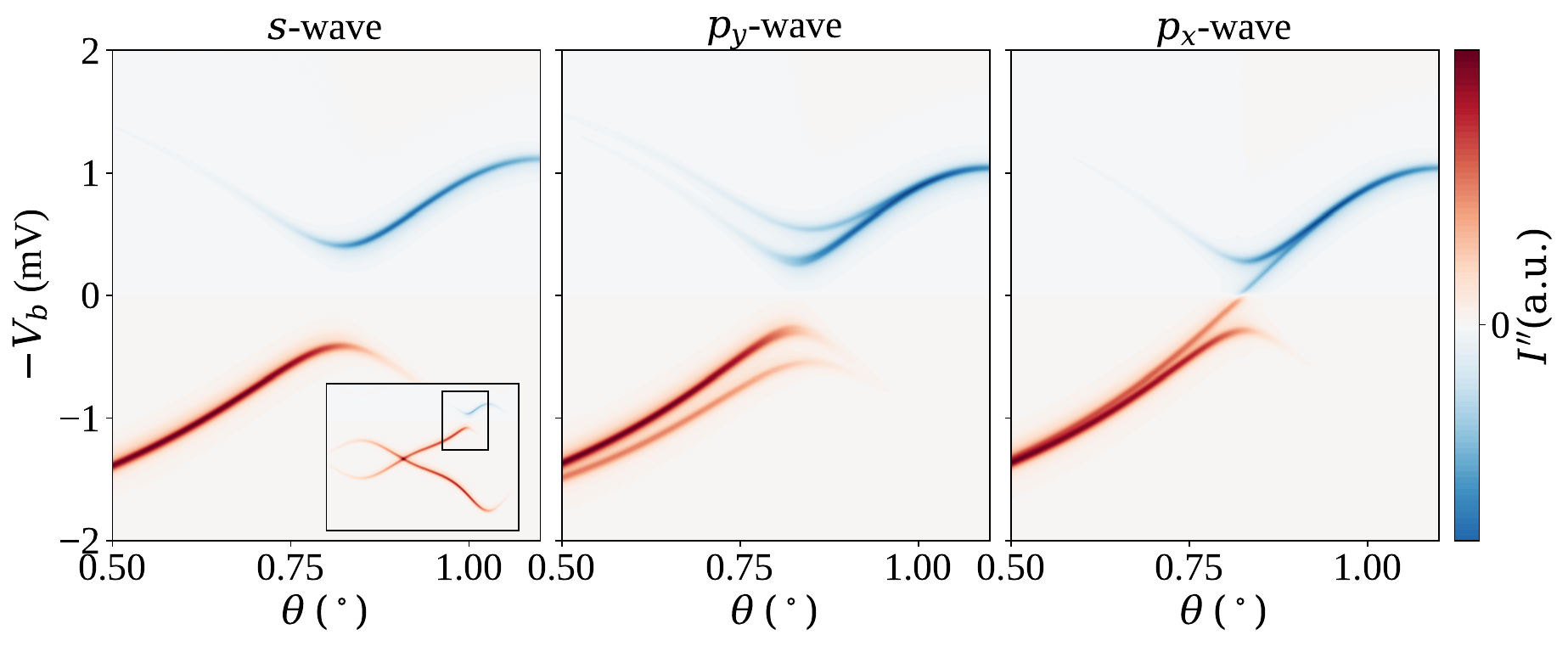}};
      \node[anchor=north west, xshift=8pt, yshift=8pt]  
        at (img.north west) {\textbf{(d)}}; 
      \node[anchor=north west, xshift=\columnwidth/3-1pt, yshift=8pt]  
        at (img.north west) {\textbf{(e)}};
      \node[anchor=north west, xshift=\columnwidth*2/3-10pt, yshift=8pt]  
        at (img.north west) {\textbf{(f)}}; 
    \end{tikzpicture}%
  \subfloat{\label{fig:deltas}}\hspace{0pt}%
  \subfloat{\label{fig:deltapx}}\hspace{0pt}%
  \subfloat{\label{fig:deltapy}}\hspace{0pt}%
  \caption{(a–c) Superconducting pairing magnitude in the mBZ for $s$-, $p_y$-, and $p_x$-wave pairings, each with three $120^\circ$-rotated line scans; the small hole pockets around $\gamma$ are shown. The high symmetry points $\kappa$, $\kappa'$, and $\gamma$ are marked in (a).
(d–f) Corresponding $I''$ spectra for BM bands with superconducting pairing. Panels show a zoom near the sample Fermi surface (inset in (d)). For $s$-wave, the three traces coincide, while for $p_{x}$ and $p_y$ pairings, broken $C_{3z}$ symmetry splits them into two distinct traces. In (f), the positive and negative branches meet at $V_b=0$, closing one gap and producing a nodal point. Parameters: $T=0.2~\si{K}$, $\Delta_0=0.4~\mathrm{meV}$, $\Gamma_{\mathrm{SC}}=0.04~\mathrm{meV}$, pairings defined in \cref{eq:pairings} in Ref.~\cite{SM}.}
\label{fig:deltascan}
\end{figure}
As a guiding example, we study the measurement of the superconducting pairing magnitude in MATBG. 
All calculations in this work are performed numerically using two-dimensional integrations over momentum space 
(see~\cite{SM} for details). We include a finite quasiparticle lifetime in the superconducting spectral function, parametrized by $\Gamma_{\mathrm{SC}}$, analogous to the Dynes broadening~\cite{dynes} (see~\cref{eq:spec_lorentzian} in Ref.~\cite{SM}).
To obtain the tunneling matrix elements and normal-state energies, we employ the continuum Bistritzer-MacDonald (BM) model~\cite{bistritzer2011moire}. 
For the rotation angle and hopping parameters, we take $\theta_{\mathrm{TBG}} = 1.1^{\circ}$, $w_{aa} = 66~\meV$, and $w_{ab} = 110~\meV$. 
We consider a state with electron doping for which the chemical potential is positive, namely $\mu_{S} = 2.5\,\text{meV}$. At this chemical potential, the Fermi surface forms a closed contour that encircles the $\gamma$ point, and the encircled area is composed of empty states (see \cref{fig:mbzs}). However, its precise geometry is not important for our analysis. We assume inter-valley pairing and a normal state that respects TRS. 
In \cref{fig:deltas}, we show that the QTM traces the energies of the Bogoliubov excitations. The minimum of this excitation spectrum yields the same gap magnitude that a conventional scanning tunneling microscope (STM) would extract, but the QTM measures the gap at a specific momentum. 

Importantly, for an MLG tip probing a graphene-based sample, tunneling occurs predominantly between the tip and the top graphene layer of the sample, and is restricted to the nearest Umklapp processes due to the exponential suppression of the tunneling amplitude at large momentum transfer~\cite{bistritzer2010transport}. Consequently, the tip Dirac point couples to three sample momenta related by $C_{3z}$ rotations around the Dirac point of the sample’s top layer, $\mathbf{K}_T$~\cite{inbar2023quantum,diracscan}. Within small-angle approximation, these momenta trace the following trajectories in TBG, defined relative to tip and sample shared $\Gamma$ point,
\begin{equation}
    \mathbf{K}_{\theta,n} = \tau \mathbf{K}_T + \tau \mathbf{q}_n \frac{\theta}{\theta_{\mathrm{TBG}}}, 
    \qquad n = 1, 2, 3,
    \label{eq:dirac_tip}
\end{equation}
where $\theta_{\mathrm{TBG}}$ is the TBG twist angle, $\tau = \pm 1$ is the valley index, $\mathbf{q}_n$ are the moiré reciprocal vectors of TBG, and $\theta$ is the rotation angle of the tip relative to the top layer of the sample.

The three tunneling channels arise from distinct Umklapp processes and add without interference at small incommensurate angles (see discussion around~\cref{eq:tun_amp_def} in Ref.~\cite{SM}, and Refs.~\cite{inbar2023quantum,diracscan}), so the measured current (or its derivatives) is the sum of three independent contributions. Each line scan thus yields three gap values, one for each $\mathbf{K}_{\theta,n}$. If the superconducting order parameter preserves $C_{3z}$ symmetry, the traces coincide for all $\theta$ (\cref{fig:mbzs,fig:deltas}), while a splitting indicates $2\pi/3$-rotational-symmetry breaking.
This behavior is shown in \cref{fig:mbzpx,fig:mbzpy}: for $p_y$- and $p_x$-wave pairings relative to the $\gamma$ point, the spectra in \cref{fig:deltapy,fig:deltapx} reveal two gaps for $p_y$ (two of the three $C_3$-related states are degenerate by mirror symmetry in $|\Delta_\mathbf{k}|$ with respect to $k_x=0$.) and a nodal point for $p_x$ where one gap vanishes (along the purple line in~\cref{fig:deltapy}). More generally, any node intersecting a trajectory produces a zero gap, enabling its direct identification. In MATBG, a nodal line along the armchair direction in momentum space crossing the $\gamma$ point would therefore be directly observable with the QTM, provided the Fermi surface encloses $\gamma$.

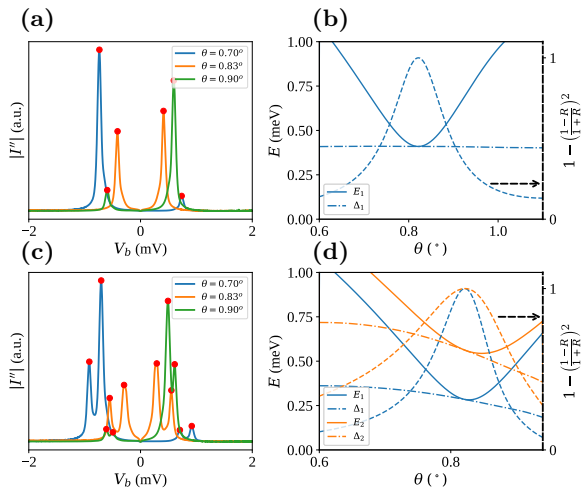
\begin{figure}[t!]
{
  \centering

    \begin{tikzpicture}
    \node[anchor=south west, inner sep=0] (img) at (0,0)
      {\includegraphics[width=\columnwidth]{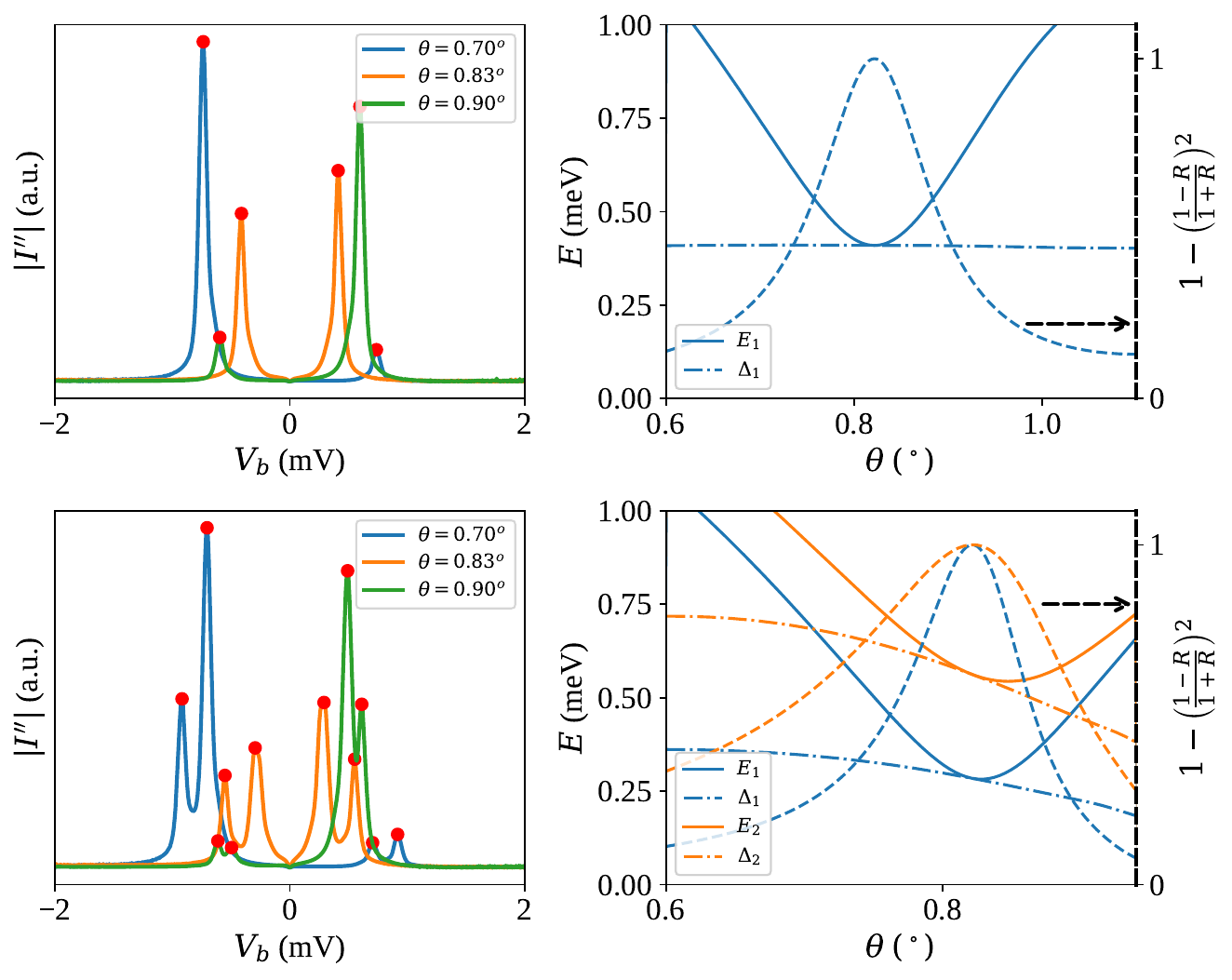}};
    \begin{scope}[
      x={(img.south east)-(img.south west)},
      y={(img.north west)-(img.south west)}
    ]
      \def\xoff{0.06} 
      \def\yoff{-0.02} 
      \node at ({\xoff},         {1-\yoff}) {\textbf{(a)}}; 
      \node at ({0.5+\xoff},     {1-\yoff}) {\textbf{(b)}}; 
      \node at ({\xoff},         {0.5-\yoff}) {\textbf{(c)}}; 
      \node at ({0.5+\xoff},     {0.5-\yoff}) {\textbf{(d)}}; 
    \end{scope}
    \end{tikzpicture}%
  \subfloat{\label{fig:dI2s}}\hspace{0pt}%
  \subfloat{\label{fig:cohrences}}\hspace{0pt}%
  \subfloat{\label{fig:dI2px}}\hspace{0pt}%
\subfloat{\label{fig:coherencepx}}\hspace{0pt}%
  }%
  \caption{(a) Cuts of $|I''|$ for $s$-wave pairing at selected twist angles $\theta$, showing coherence peaks whose intensities are proportional to the Bogoliubov coherence factors (marked by red dots). 
  (b) Bogoliubov excitations dispersions $E_{\mathbf{k}}$ extracted from the spectra in (a), together with the pairing magnitude $\Delta$ obtained from the combination of $E_{\mathbf{k}}$ and the peak intensity ratio, $R$, see discussion around~\cref{eq:R}. 
  (c) Same as (a) but for $p_{y}$-wave pairing, where two pairs of symmetric peaks are resolved.
  (d) Quasiparticle dispersions and extracted pairing for the $p_{y}$ pairing, illustrating how the pairing magnitude varies with $\theta$ (corresponding to momentum $\kth$). The extracted pairing magnitude changes by approximately $50\%$ along the trajectory.}
  \label{fig:coherence}
\end{figure}
As seen in~\cref{eq:tun_sc}, the QTM’s in-plane momentum conservation allows the Bogoliubov coherence factors $u_{\mathbf{k}}$ and $v_{\mathbf{k}}$ to be probed for the three lines of momenta. This is in contrast to STM measurements, which integrate over in-plane momentum, such that the measured conductance reflects only the density of states and averages out the momentum dependence of the coherence factors. As the tip Dirac point traces $\kth$ across the mBZ, the QTM resolves the quasiparticle peaks at $eV_{b\pm}=\pm E_{\mathbf{k}}$. Under the assumptions used to derive \cref{eq:tun_sc}, their contributions scale as $I''_+(\theta) \propto |T(\kth)|^{2}|u_{\kth}|^{2}$ for $+E_{\kth}$ and as $I''_-(\theta)\propto |T(\kth)|^{2}|v_{\kth}|^{2}$ for $-E_{\kth}$. 
Because both share the same tunneling matrix element, it cancels out in the ratio, which becomes
\begin{equation}
R(\kth) 
\equiv 
\frac{I''_{+}(\theta)}{I''_{-}(\theta)}
= \frac{|u_{\kth}|^{2}}{|v_{\kth}|^{2}}
= \frac{E_{\kth}+\xi_{\kth}}
       {E_{\kth}-\xi_{\kth}}.
     \label{eq:R}
\end{equation}
The energy separation between the positive and negative voltage peaks at the same momentum fixes $E_{\mathbf{k}}$ through $e(V_{b+}-V_{b-})=2E_{\mathbf{k}}$, while the ratio of their intensities yields $R(\mathbf{k})$. Together, they give 
\begin{equation}
\left|\Delta_{\kth}\right|=E_{\kth}\sqrt{1-\left(\frac{1-R(\kth)}{1+R(\kth)}\right)^{2}},
\label{eq:coh_extract}
\end{equation}
offering a novel approach to directly extract the pairing magnitude at each probed momentum, even away from the Fermi momentum.

\cref{fig:coherence} illustrates this procedure. For an $s$-wave state, the $I''$ spectra (\cref{fig:dI2s}) show  a pair of bias-symmetric coherence peaks whose height ratio yields $|u_{\mathbf{k}}|^{2}/|v_{\mathbf{k}}|^{2}$, and the reconstructed pairing magnitude is isotropic (\cref{fig:cohrences}). In contrast, for $p_{y}$ symmetry, for each $\theta$, two distinct pairs of peaks appear along the bias scan (\cref{fig:dI2px}), and the extracted pairing varies with the twist angle $\theta$ (corresponding to momentum $\kth$), (\cref{fig:coherencepx}). We note that a similar idea was proposed for angle-resolved photoemission spectroscopy (ARPES) measurements~\cite{bogoliubov_angle1,bogoliubov_angle2}; however, whereas ARPES can access the positive energy branch only at high temperatures when these states are partially occupied, the QTM can do so even at low temperatures.

In some systems, superconductivity might be based on a valley-polarized normal state that breaks TRS~\cite{liu2022isospin,kim2025topological,chou2025intravalley,parra2025band,han2025signatures,may2025pairing,friedlan2025valley,sedov2025probing,christos2023nodal}. Therefore, the relation $\xi_\mathbf{k}=\xi_{-\mathbf{k}}$ does not necessarily hold. However, the excitation energies are only slightly modified, $E_\mathbf{k}=\delta \xi_\mathbf{k}\pm \sqrt{\bar{\xi}_\mathbf{k}^2+\Delta_\mathbf{k}^2}$, where $\delta\xi_\mathbf{k}=\frac{\xi_\mathbf{k}-\xi_{-\mathbf{k}}}{2}$ and $\bar{\xi}_\mathbf{k}=\frac{\xi_\mathbf{k}+\xi_{-\mathbf{k}}}{2}$. The coherence factors in \cref{eq:coherenc} remain the same, up to the change $\xi_{\mathbf{k}}\rightarrow\bar{\xi}_{\mathbf{k}}$. \cref{eq:coh_extract} remains valid; only $E_\mathbf{k}$ is now measured relative to $\delta\xi_{\mathbf{k}}$ and the peaks are no longer symmetric around zero bias.

In the measurement configuration discussed so far, scanning with a Dirac tip probes tunneling only in the vicinity of the tip Dirac point. 
Consequently, the accessible information is restricted to a narrow region of momentum space, providing only partial insight into the superconducting pairing structure and leaving possible nodal points along the Fermi surface out of reach. 

We demonstrate that operating in a different modality allows the measurement to be extended to a significantly larger portion of momentum space, and in particular enables the identification of nodal points in the pairing potential, if they exist. 
To this end, we consider the case where the tip has a non-zero density, and therefore a Fermi surface, and focus on zero bias voltage, $V_b = 0$. 
Experimentally, this is achieved by tuning the gate voltages so that the tip chemical potential $\mu_T$ and the electrostatic shift $\phi$ vary simultaneously while maintaining $V_b \!\approx\! 0$. 
This configuration enables the measurement to probe quasiparticle excitations at the center of the superconducting gap, i.e., around $\mu_S$.

In this regime, the zero-bias conductance ($I'$) exhibits a peak whenever the circular Fermi surface of the tip—with radius 
$k_F^{T} = |\mu_T| / (\hbar v_D)$ 
and centered at the tip Dirac momentum $\mathbf{K}_\theta$—crosses a nodal point $\mathbf{k}_0$. 
This scheme therefore provides a direct method to determine the radial momentum distance between a well-defined reference point in momentum space (the tip Dirac point) and a nodal point. 
Repeating the measurement for two different tip rotation angles, $\theta_1$ and $\theta_2$, corresponding to Fermi circle centers $\mathbf{K}_{\theta_1}$ and $\mathbf{K}_{\theta_2}$, makes it possible to triangulate the nodal momenta geometrically from the intersection of the two tip Fermi circles (see~\cref{fig:nodal_loc} in Ref.~\cite{SM}).

\begin{figure}[t]{
    \centering
    \begin{tikzpicture}
    \node[anchor=south west, inner sep=0] (img) at (0,0){
  \includegraphics[width=\columnwidth]{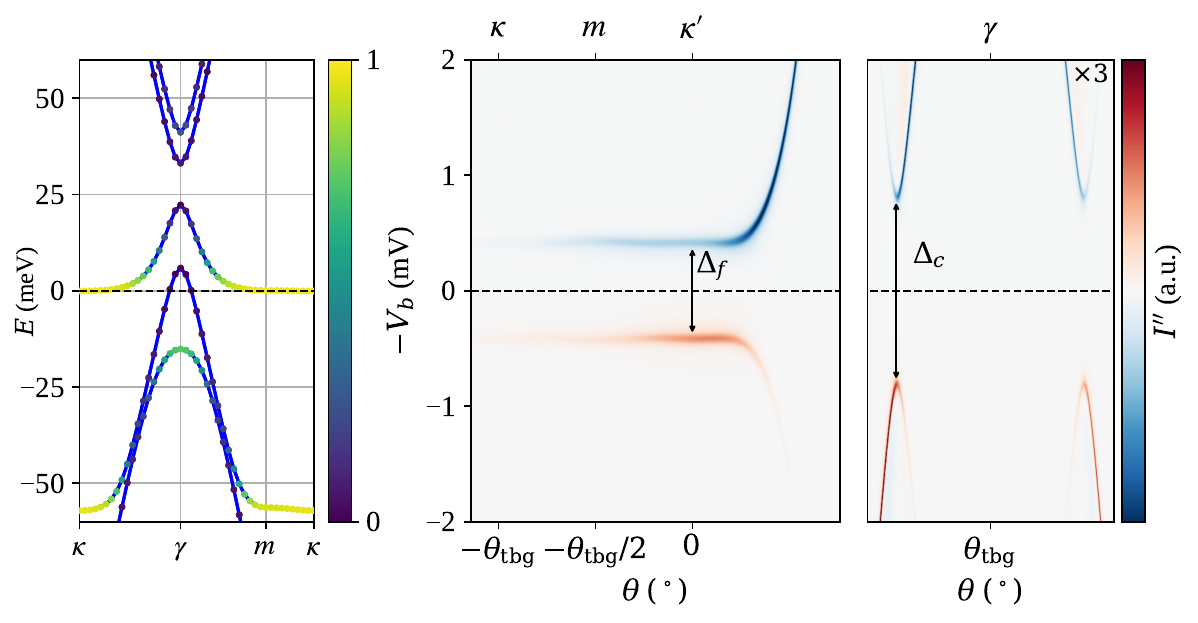}};
    \begin{scope}[
      x={(img.south east)-(img.south west)},
      y={(img.north west)-(img.south west)}
    ]
      \def\xoff{0.04} 
      \def\yoff{0.02} 
      \node at ({0.04+\xoff},         {1-\yoff}) {\textbf{(a)}}; 
      \node at ({0.32+\xoff},     {1-\yoff}) {\textbf{(b)}}; 
      \node at ({0.7+\xoff},     {1-\yoff}) {\textbf{(c)}}; 
    \end{scope}
    \end{tikzpicture}%
  \subfloat{\label{fig:bands_hf}}\hspace{0pt}%
  \subfloat{\label{fig:sc_hf1}}\hspace{0pt}%
 \subfloat{\label{fig:sc_hf2}}\hspace{0pt}%
    }
    \caption{(a) MATBG band structure at $\nu=-2$ from the heavy-fermion model using a one-shot mean-field and K-IVC parent state. Color denotes $f$-electron weight. The flat, chiral limit is taken with $v'_\star=0$ and $M=0$. The dashed line marks a possible Fermi level assuming weak doping that does not modify the bands.
(b,c) QTM $I''$ spectra for superconducting states. (b) Fermi level in the flat $f$ band opens gaps near $\kappa$, $m$, and $\kappa'$ (tip rotations $-\theta_{\mathrm{TBG}}$, $-\theta_{\mathrm{TBG}}/2$, $0$); (c) Fermi level in the dispersive $c$ band opens a gap around $\gamma$ ($\theta_{\mathrm{TBG}}$). Parameters: $w_{aa}/w_{ab}=0.8$, $\theta_{\mathrm{TBG}}=1.05^\circ$, $\Gamma_{\mathrm{SC}}=0.04~\si{meV}$, $T=0.2~\si{K}$, $\Delta_f=0.4~\si{meV}$, $\Delta_c=0.8~\si{meV}$. Color intensity near $\gamma$ is scaled by $3$ to account for degeneracy.}
\label{fig:fig4}
\end{figure}

Up to this point, we have analyzed the QTM spectrum using a noninteracting quadratic model. To include interaction effects in twisted flat-band systems, we employ the heavy-fermion (HF) framework~\cite{thfsong},  which has been demonstrated to successfully capture the correlated band structure of MATBG~\cite{xiao2025interacting}. The model, consisting of localized $f$ electrons hybridizing with itinerant Dirac-like $c$ electrons, extends the noninteracting description in a one-shot mean-field approximation that captures strong correlations.

Without interactions, the $f$ and $c$ states hybridize and recover the BM band structure. Ref.~\cite{thfsong} derives the interacting Hamiltonian in the mean-field approximation and shows that evaluating it with a parent state of occupied $f$ orbitals (one-shot) yields results that closely match the fully self-consistent solution at integer filling.

Following this one-shot mean-field approach, we calculate the band structure and tunneling matrix elements in the HF basis. \cref{fig:bands_hf} shows the resulting band structure for the $\nu=-2$, Kramers intervalley coherent (K-IVC) state~\cite{bultinck2020ground}, obtained under the simplifying assumptions of chiral symmetry (see Ref.~\cite{SM} for details). We further examine a possible Fermi-level alignment near charge neutrality, assuming that small doping does not significantly modify the band structure. In this configuration, the Fermi energy intersects the flat $f$ bands, producing a Fermi surface near the edges of the mBZ, and also crosses the dispersive $c$ bands, resulting in a small Fermi pocket around the $\gamma$ point. The exact topology of the Fermi surface in MATBG remains uncertain~\cite{kang2021cascades,datta2023heavy,rai2024dynamical,zhou2024kondo,calderon2025cascades,merino2025interplay,cualuguaru2025obtaining}; this alignment is considered because it includes two distinct types of Fermi surfaces that may occur separately or coexist.

We consider isotropic superconducting pairing and calculate the QTM spectra for a pairing located at the edge of the mBZ (\cref{fig:sc_hf1}) and at the center of the mBZ (\cref{fig:sc_hf2}). This comparison highlights a key advantage of the QTM. Whereas STM would detect a gapped spectrum in both cases, the QTM can resolve the specific tip rotation angles at which the gap appears. A gap observed at $-\theta_{\mathrm{TBG}} < \theta < 0$ indicates pairing predominantly among flat $f$-electron states, while a gap appearing near $\theta_{\mathrm{TBG}}$, associated with the dispersive band, signifies pairing among $c$ electrons. Thus, the QTM provides a direct probe to address the open question of whether superconducting pairing in MATBG arises from $f$ or $c$ components, as well as the nature of the quasiparticles around the Fermi energy. For simplicity, we assume isotropic pairing between states related by spinless time-reversal symmetry~\cite{bultinck2020ground} and neglect any spin or valley dependence. Our previous analysis regarding the detection of $C_{3z}$-symmetry breaking, coherence-factor ratios, and nodal point detection remains valid for the mean-field HF, since this Hamiltonian yields well-defined quasiparticle eigenstates, similar to the BM model.

\emph{Discussion and Conclusions} We have shown that the quantum twisting microscope (QTM) can be used as a momentum-resolved probe of superconductivity.
By rotating the QTM tip in real space, the tip Dirac point maps the Bogoliubov excitations $E_{\mathbf{k}}$ along well-defined trajectories in momentum space.
The relative intensities of the peaks in $I''$ correspond to electron and hole excitations, allowing one to directly extract the coherence factor magnitudes $|u_{\mathbf{k}}|^{2}$ and $|v_{\mathbf{k}}|^{2}$, and hence determine the pairing amplitude $|\Delta_{\mathbf{k}}|$, even away from the Fermi momentum $k_F$.
Because elastic tunneling originates from three Umklapp scattering processes related by $C_{3z}$ rotations, a single line scan yields three pairing measurements. Their degeneracy tests $C_{3z}$ symmetry, and any splitting that appears only in the superconducting state points to nematicity in the pairing. A vanishing gap identifies a nodal point on the line scan. 
A complementary, zero-bias measurement that expands the tip Fermi circle produces sharp zero-bias resonances when the circle crosses a node, enabling geometric triangulation of the nodal momenta. The same protocols can be applied to MATBG and related moiré superconductors, and can be combined with normal-state measurements for quantitative calibration of tunneling matrix elements and for identifying the nature of the normal state. Within the heavy-fermion description of MATBG, the QTM can further identify whether superconducting pairing arises from the flat ($f$) or dispersive ($c$) bands.

In the weak-tunneling regime considered here, the QTM current is weighted by a tunneling matrix element that predominantly projects onto the sample top layer and can be calibrated from tunneling measurements in the normal state. Deviations from the expected matrix elements beyond Bogoliubov coherence-factor weights may therefore indicate a redistribution of eigenstates weight among different layer or sublattice components.

In general, the superconducting order parameter is a matrix in spin and valley space. The present work focuses on unitary pairing, for which the quasiparticle spectrum is characterized by a single scalar pairing magnitude. Because the QTM probes the single-particle spectral function, it is sensitive to the eigenvalues of the pairing matrix magnitude ($\Delta^\dagger_{\mathbf{k}}\Delta_{\mathbf{k}}$), and non-unitary pairing matrices can manifest as splittings of the Bogoliubov quasiparticle branches or the appearance of multiple gaps. In the current implementation with a graphene tip, the QTM current conserves spin and valley and sums over them with equal weight, so their structure cannot be directly resolved. Introducing spin or valley selectivity, for example via proximity-induced spin–orbit coupling in the tip or an in-plane magnetic field, could provide additional insight into the internal flavor structure of the pairing.
Looking ahead, extending to stronger tunneling regimes~\cite{andrev} could enable phase-sensitive and momentum-conserving Andreev spectroscopy.

Overall, the QTM provides a powerful, momentum-resolved platform to measure the pairing potential and characterize unconventional superconductivity in two-dimensional and moiré materials.

\emph{Note added} -- While preparing our manuscript for publication, we learned of a related study~\cite{wei2025superconductors}, conducted independently.
\begin{acknowledgments}
We thank Shahal Ilani, Felix von Oppen, Moran Shapira, Yaar Vitury, and Jiewen Xiao for helpful discussions.
This work was supported by the Israeli Science Foundation, the Israeli Ministry of Science Technology and Space, the Minerva Stiftung, the
DFG (CRC/Transregio 183, EI 519/7-1), the Israel Science Foundation ISF (Grant No 1914/24) and ISF Quantum Science and Technology (2074/19).
\end{acknowledgments}
\bibliography{refs}

\clearpage
\onecolumngrid
\begin{center}
\textbf{Supplementary Materials}
\end{center}
\renewcommand{\thefigure}{S\arabic{figure}}
\renewcommand{\thetable}{S\arabic{table}}
\renewcommand{\thesection}{S\arabic{section}}
\renewcommand{\theequation}{S\arabic{equation}}
\setcounter{equation}{0}
\setcounter{figure}{0}
\tableofcontents
\clearpage

\section{Electrostatics of the QTM junction}
\label{app:electrostatics}
We follow Ref.~\cite{theoryphonons} to derive the electrostatic relations governing the QTM junction, with minor modifications.
For the tip (T) and sample (S), we define the electrochemical potential 
$\mu_X^{\mathrm{ec}}=\mu_X-\phi_X$ with $X\in\{T,S\}$, 
where $\mu_X$ is the intrinsic chemical potential and $\phi_X$ is the electrostatic potential. 
A bias voltage $V_b$ determines their difference as
\begin{equation}
\mu_T^{\mathrm{ec}}-\mu_S^{\mathrm{ec}}=-eV_b 
\;\;\Longleftrightarrow\;\;
-eV_b = \mu_T - \mu_S - \phi_T+\phi_S,
\end{equation}
where $\phi=\phi_S-\phi_T$ is the electrostatic potential difference across the junction (in units of energy).  

We consider the circuit illustrated in \cref{fig:QTM_a} in the main text, where the tip, sample, and gates are modeled as a series of infinite parallel-plate capacitors.  
Assuming a symmetric bias configuration, the potentials satisfy
\begin{equation}
\frac{eV_b}{2} =\mu_{S}- \phi_{S} ,\quad -\frac{eV_b}{2}=\mu_T-\phi_T.
\end{equation}
The capacitor equations for symmetric top and bottom gates are
\begin{equation}
\begin{aligned}
&(\phi_{\mathrm{TG}}-\phi_T) = -\frac{e^{2}}{\varepsilon_0 \varepsilon_{\mathrm{g}}}\, d_g\, n_{\mathrm{TG}}, \\
&(\phi_T-\phi_S) = -\frac{e^{2}}{\varepsilon_0 \varepsilon}\, d\, (n_T+n_{\mathrm{TG}})
                 = \frac{e^{2}}{\varepsilon_0 \varepsilon}\, d\, (n_S+n_{\mathrm{BG}}), \\
&(\phi_S-\phi_{\mathrm{BG}}) = \frac{e^{2}}{\varepsilon_0 \varepsilon_{\mathrm{g}}}\, d_g\, n_{\mathrm{BG}},
\label{eq:capacitance}
\end{aligned}
\end{equation}
where $\phi_{\mathrm{TG}},\phi_T,\phi_S,\phi_{\mathrm{BG}}$ are the electrostatic potentials of the top gate (TG), tip, sample, and bottom gate (BG).  
$n_{\mathrm{TG}}, n_{\mathrm{BG}}, n_T,$ and $n_S$ are the corresponding charge densities. $d_g$ and $\varepsilon_g$ are the dielectric thickness and relative permittivity of the layers separating the top gate from the tip and the bottom gate from the sample, 
while $d$ and $\varepsilon$ denote the dielectric thickness and relative permittivity 
of the layer between the sample and the tip. $\varepsilon_0$ is the vacuum permittivity.

We note that \cref{eq:capacitance} represents an effective capacitor model for the junction and neglects screening effects arising from the out-of-plane polarization of the tip and sample two-dimensional sheets. To incorporate this effect, it is common to model each sheet as having an effective dielectric thickness $\ell$, which reduces the electrostatic potential drop between the layers by an amount proportional to $\ell$ and to the average displacement field immediately above and below the sheet~\cite{slizovskiy2021out,boschi2024built}. Denoting $\ell_T$ and $\ell_S$ as the dielectric thicknesses of the tip and sample, respectively, \cref{eq:capacitance} is modified to
\begin{equation}
\begin{aligned}
&(\phi_{\mathrm{TG}}-\phi_T)
= -\frac{e^{2}}{\varepsilon_0 \varepsilon_{\mathrm{g}}}\, d_g\, n_{\mathrm{TG}}+\frac{e^{2}\,\ell_T}{4\varepsilon_0}\left(\frac{n_{\mathrm{TG}}}{\varepsilon_{\mathrm g}}+\frac{n_T+n_{\mathrm{TG}}}{\varepsilon}\right),
\\
&(\phi_T-\phi_S)= -\frac{e^{2}}{\varepsilon_0 \varepsilon}\left(d-\frac{\ell_T+\ell_S}{4}\right)(n_T+n_{\mathrm{TG}})+\frac{e^{2}}{4\varepsilon_0\varepsilon_{\mathrm g}}\left(\ell_T n_{\mathrm{TG}}-\ell_S n_{\mathrm{BG}}\right)
\\
&\hspace{1.8cm}= \frac{e^{2}}{\varepsilon_0 \varepsilon}\left(d-\frac{\ell_T+\ell_S}{4}\right)(n_S+n_{\mathrm{BG}})+\frac{e^{2}}{4\varepsilon_0\varepsilon_{\mathrm g}}\left(\ell_T n_{\mathrm{TG}}-\ell_S n_{\mathrm{BG}}\right),
\\
&(\phi_S-\phi_{\mathrm{BG}})= \frac{e^{2}}{\varepsilon_0\varepsilon_{\mathrm{g}}}\, d_g\, n_{\mathrm{BG}}-\frac{e^{2}\,\ell_S}{4\varepsilon_0}\left(\frac{n_S+n_{\mathrm{BG}}}{\varepsilon}+\frac{n_{\mathrm{BG}}}{\varepsilon_{\mathrm g}}\right).
\end{aligned}
\end{equation}
For graphene layers, the dielectric thickness is $\ell_T=\ell_S=\ell \approx 2.1\,\si{\angstrom}$~\cite{slizovskiy2021out,boschi2024built}. In the following analysis, we neglect the dielectric thickness to simplify the qualitative discussion, motivated by the assumption that the dielectric thickness of the tunneling barrier is larger than $\ell$ and dominates the capacitance. Nevertheless, a quantitatively accurate treatment of the electrostatics should take this effect into account.

Assuming metallic gates with a large density of states, their applied voltages fix the potentials such that $eV_{\mathrm{TG/BG}}=\phi_{\mathrm{TG/BG}}$.  
We define the specific capacitances:
\begin{equation}
C=\frac{\varepsilon_0\varepsilon}{d},\quad C_g=\frac{\varepsilon_0\varepsilon_g}{d_g}.
\end{equation}
So \cref{eq:capacitance} can be reduced to three compact relations:
\begin{equation}
\begin{aligned}
e(V_{\mathrm{TG}}-V_{\mathrm{BG}}) &= \frac{e^{2}}{C_g}( n_T- n_S)-\left(1+\frac{2C}{C_g}\right)\phi,\\
e(V_{\mathrm{TG}}+V_{\mathrm{BG}}) &= \frac{e^{2}}{C_g}( n_T+ n_S)+\mu_T+\mu_S,\\
-eV_b &= \phi + \mu_T - \mu_S.
\label{eq:electrostat}
\end{aligned}
\end{equation}
\cref{eq:electrostat} determines the three unknowns $\{\phi,\mu_T,\mu_S\}$ for a given device geometry, gates, and bias voltages.
We first consider the limit of large gate capacitance ($C_g \gg C$, or equivalently $d_g \ll d$). 
This limit poses experimental challenges: increasing the thickness of the tunneling barrier significantly reduces the current across the junction, 
while reducing the gate dielectric thickness may lead to leakage currents. 
Nevertheless, analyzing the electrostatic equations in this limit provides useful insight into the required gate-voltage compensation. 
In this regime, the equations simplify to
\begin{equation}
\begin{aligned}
eV_{\mathrm{TG}} - \tfrac{eV_b}{2} &= \frac{e^{2}}{C_g}\, n_T(\mu_T) + \mu_T,\\
eV_{\mathrm{BG}} + \tfrac{eV_b}{2} &= \frac{e^{2}}{C_g}\, n_S(\mu_S) + \mu_S,\\
-eV_b &= \phi + \mu_T - \mu_S .
\end{aligned}
\end{equation}
These relations show that in this limit $V_{\mathrm{TG}} - V_b/2$ primarily controls the tip doping, while $V_{\mathrm{BG}} + V_b/2$ controls the sample doping. 
Thus, changing $V_b$ while compensating $V_{\mathrm{TG}}$ and $V_{\mathrm{BG}}$ allows one to vary $\phi$ without changing $\mu_T$ and $\mu_S$.
Returning to the general case of \cref{eq:electrostat} (without assuming $C_g\!\gg\!C$), we derive the conditions required to vary each of $\phi$, $\mu_T$, and $\mu_S$ independently.  
Defining
\begin{equation}
V_G=\tfrac{1}{2}(V_{\mathrm{TG}}+V_{\mathrm{BG}}),\quad 
V_D=\tfrac{1}{2}(V_{\mathrm{TG}}-V_{\mathrm{BG}}),
\end{equation}
and differentiating \cref{eq:electrostat} with respect to $\phi$, while keeping $\mu_T$ and $\mu_S$  constant (enforcing $\frac{\partial\mu_T}{\partial\phi}=\frac{\partial\mu_S}{\partial\phi}=0$), gives
\begin{equation}
\begin{aligned}
e\,\frac{\partial V_G}{\partial \phi} &= 0, \\
e\,\frac{\partial V_D}{\partial \phi} &= -\left(\frac12 + \,\frac{C}{C_g}\right), \\
e\,\frac{\partial V_b}{\partial \phi} &= -1.
\label{eq:constphi}
\end{aligned}
\end{equation}
Therefore, a change in $\phi$ can be induced directly via the bias, $e\,\Delta V_b =- \Delta \phi$, with gate compensation given by $\Delta V_G = 0$ and $e\,\Delta V_D = (\frac{1}{2} + \frac{C}{C_g
})\,\Delta V_b$.  
Experimentally, $C$ and $C_g$ can be determined from fits to data~\cite{inbar2023quantum,xiao2025interacting,qtmphonons}. If the tunneling barrier and the barrier between the junction and the gates are of the same material, the capacitance ratio ${C}/{C_g}$ can be calculated using the ratio of their thickness.

Differentiating \cref{eq:electrostat} with respect to $\mu_T$ and $\mu_S$ yields
\begin{equation}
\begin{aligned}
e\,\frac{\partial V_b}{\partial \mu_T} &= -1, \\
e\,\frac{\partial V_{\mathrm{TG}}}{\partial \mu_T} &= \frac{e^{2}}{C_g}\,\frac{\partial n_T}{\partial \mu_T} + \tfrac{1}{2}, \\
e\,\frac{\partial V_{\mathrm{BG}}}{\partial \mu_T} &= \tfrac{1}{2},
\end{aligned}
\end{equation}
and
\begin{equation}
\begin{aligned}
e\,\frac{\partial V_b}{\partial \mu_S} &= 1, \\
e\,\frac{\partial V_{\mathrm{TG}}}{\partial \mu_S} &= \tfrac{1}{2}, \\
e\,\frac{\partial V_{\mathrm{BG}}}{\partial \mu_S} &= \frac{e^{2}}{C_g}\,\frac{\partial n_S}{\partial \mu_S} + \tfrac{1}{2},
\end{aligned}
\end{equation}
respectively. Hence, controlling $\mu_T$ independently requires knowledge of the tip density of states, namely, $\partial n_T/\partial \mu_T$.  
For a graphene tip, this dependence is well known, and only the absolute doping needs to be calibrated.  
Since $\mu_T=0$ corresponds to the Dirac point where the tip's density of states vanishes (yielding zero tunneling current), this calibration can often be performed experimentally~\cite{xiao2025interacting}.  
In contrast, the density of states of the sample, $n_S(\mu_S)$, may have a more complex dependence, making it necessary to determine the sample doping independently using complementary probes.

\section{Calculation of the tunneling current and tunneling matrix element}
\label{app:tunnel_matrix}

Here, we briefly review the tunneling current and tunneling matrix element in the QTM with a graphene tip and a sample whose top layer is graphene, following previous works~\cite{bistritzer2010transport,inbar2023quantum,diracscan}.  

In the weak-tunneling regime, the QTM current is evaluated within the tunneling Hamiltonian approximation~\cite{mahan1990many},
\begin{equation}
I(V_b) = \frac{2\pi e}{\hbar} 
\sum_{\mathbf{k}_T,\mathbf{k}_S,\alpha,\beta}
\left|T_{\alpha\beta}(\mathbf{k}_T,\mathbf{k}_S)\right|^2 
\int d\omega \,
A_{T,\alpha}(\mathbf{k}_T,\omega+eV_b) A_{S,\beta}(\mathbf{k}_S,\omega) \left[ f(\omega) - f(\omega+eV_b) \right] ,
\label{eq:tun_cur}
\end{equation}
where $e$ is the elementary charge, $\mathbf{k}_T$ and $\mathbf{k}_S$ are the Bloch-state momenta, $T_{\alpha\beta}(\mathbf{k}_T,\mathbf{k}_S)$ is the tunneling matrix element between the tip ($T$) and sample ($S$), $\alpha$ and $\beta$ label the band indices, $A_{T,S}(\mathbf{k},\omega)$ are the spectral functions of the tip and sample, and $f(\omega)$ is the Fermi function. The summation over spin and valley degrees of freedom is implicit.

We assume that the tip lifetime broadening is negligible compared to the sample broadening, so the tip spectral function is taken as $A_{T,\alpha}(\omega,\mathbf{k})=\delta(\omega-\xi_{\mathbf{k},\alpha})$, where $\alpha$ labels the tip bands and $\xi_{\mathbf{k},\alpha}$ is the dispersion measured relative to the tip chemical potential. Under this assumption, the integral over $d\omega$ can be evaluated, yielding  
\begin{equation}
    I = \frac{2\pi e}{\hbar} 
\sum_{\mathbf{k}_T,\mathbf{k}_S,\alpha,\beta}
\left|T_{\alpha\beta}(\mathbf{k}_T,\mathbf{k}_S)\right|^2
A_{S,\beta}(\mathbf{k}_S,\xi_{\mathbf{k}_T,\alpha}-eV_b) 
\left[ f(\xi_{\mathbf{k}_T,\alpha}-eV_b) - f(\xi_{\mathbf{k}_T,\alpha}) \right].
\end{equation}
Even if the tip has a finite lifetime broadening, the convolution of two Lorentzian spectral functions remains a Lorentzian, with a width equal to the sum of the individual widths. Thus, the following analysis remains valid up to a renormalization of the effective sample lifetime.

The tunneling matrix element between the twisted graphene layers is given by~\cite{bistritzer2010transport}  
\begin{equation}
T_{\alpha\beta}(\mathbf{k}_T,\mathbf{k}_S) =
\frac{1}{\Omega_0}
\sum_{\mathbf{G}_T,\mathbf{G}_S}
t(\mathbf{k}_S + \mathbf{G}_S)
\bra{u^T_{\mathbf{k}_T,\alpha}}T_{\mathbf{G}_T,\mathbf{G}_S} \ket{u^{S,\text{top}}_{\mathbf{k}_S,\beta}}
\delta_{\mathbf{k}_T+\mathbf{G}_T,\,\mathbf{k}_S+\mathbf{G}_S},
\end{equation}
where $\Omega_0$ is the unit-cell area, $t(\mathbf{k})$ is the Fourier transform of the interlayer hopping amplitude, $\ket{u^T_{\mathbf{k}_T,\alpha}}$ is the Bloch wavefunction of the tip in the sublattice basis, and $\ket{u^{S,\text{top}}_{\mathbf{k}_S,\beta}}$ is the top-layer component of the sample Bloch wavefunction. The sum runs over reciprocal lattice vectors of the tip graphene layer ($\mathbf{G}_T$) and the top graphene layer of the sample ($\mathbf{G}_S$). 

In the sublattice basis, the matrix $T_{\mathbf{G}_T,\mathbf{G}_S}$ is given by
\begin{equation}
T_{\mathbf{G}_T,\mathbf{G}_S} =
\begin{pmatrix}
e^{i \mathbf{G}_T\cdot \boldsymbol{\tau}_T^A - i \mathbf{G}_S\cdot \boldsymbol{\tau}_S^A} & e^{i \mathbf{G}_T \cdot\boldsymbol{\tau}_T^A - i \mathbf{G}_S \cdot\boldsymbol{\tau}_S^B} \\
e^{i \mathbf{G}_T \cdot\boldsymbol{\tau}_T^B - i \mathbf{G}_S \cdot\boldsymbol{\tau}_S^A} & e^{i \mathbf{G}_T \cdot\boldsymbol{\tau}_T^B - i \mathbf{G}_S \cdot\boldsymbol{\tau}_S^B}
\end{pmatrix}
e^{\,i \mathbf{G}_T \cdot \mathbf{d}_T - i \mathbf{G}_S \cdot \mathbf{d}_S},
\end{equation}
where $\boldsymbol{\tau}^{A/B}_{T/S}$ are the positions of the A and B sublattices of the tip and the sample’s top layer, and $\mathbf{d}_T$, $\mathbf{d}_S$ are the displacement vectors of the layers.  

For incommensurate alignment, a unique pair $\{\mathbf{G}_T,\mathbf{G}_S\}$ satisfies momentum conservation for each $\{\mathbf{k}_T,\mathbf{k}_S\}$. Fixing $\mathbf{k}=\mathbf{k}_S+\mathbf{G}_S=\mathbf{k}_T+\mathbf{G}_T$ uniquely determines $\mathbf{k}_S,\mathbf{G}_S,\mathbf{k}_T,\mathbf{G}_T$. Thus, we define  
\begin{equation}
    \left.T_{\alpha\beta}(\mathbf{k})\right|_{\mathbf{k}=\mathbf{k}_S+\mathbf{G}_S=\mathbf{k}_T+\mathbf{G}_T} \equiv \frac{1}{\Omega_0}
t(\mathbf{k})
\bra{u^T_{\mathbf{k}_T,\alpha}}T_{\mathbf{G}_T,\mathbf{G}_S} \ket{u^{S,\text{top}}_{\mathbf{k}_S,\beta}}.
\label{eq:tun_amp_def}
\end{equation}
Since $t(\mathbf{k})$ decays exponentially with momentum, tunneling is restricted to the first Brillouin zone of the twisted tip–sample system~\cite{bistritzer2011moire}. Further assuming that $t(\mathbf{k})$ varies weakly in this region, we approximate $t(\mathbf{k})\approx t_0$ and write the matrices $T_{\mathbf{G}_T,\mathbf{G}_S}$ for the three allowed Umklapp processes, separating by valley $\tau=\pm1$:  
\begin{align}
T_{1,\tau} &=
\begin{pmatrix}
1 & 1 \\
1 & 1
\end{pmatrix}
, \\[1em]
T_{2,\tau} &=
\begin{pmatrix}
1 & e^{ -i \tfrac{2\pi}{3}\tau} \\
e^{i \tfrac{2\pi}{3}\tau} & 1
\end{pmatrix}, \\[1em]
T_{3,\tau} &=
\begin{pmatrix}
1 & e^{i \tfrac{2\pi}{3}\tau} \\
e^{- i \tfrac{2\pi}{3}\tau} & 1
\end{pmatrix}.
\end{align}
Although the phases $e^{i\tau( \mathbf{G}_T \cdot \mathbf{d}_T - \mathbf{G}_S \cdot \mathbf{d}_S)}$ differ between the three processes, at incommensurate angles, a given pair $(\mathbf{k}_T,\mathbf{k}_S)$ satisfies momentum conservation through at most one channel. Thus, the three processes contribute without interference, and the phases drop out when taking absolute values.  

Separating the contributions from each valley, the current becomes   
\begin{equation}
    I = \frac{2\pi  |t_0|^2e}{\Omega^2_0\hbar} 
\sum_{\tau=\pm1}\sum_{n=1}^{3}\sum_{\alpha,\beta}\sum_{\mathbf{p}}
\left|\bra{u^T_{\mathbf{p},\tau,\alpha}}T_{n,\tau} \ket{u^{S,\text{top}}_{\mathbf{p},\tau,\beta}}\right|^2
A_{S,\beta}(\mathbf{p},\xi_{\mathbf{p},\alpha}-eV_b) 
\left[ f(\xi_{\mathbf{p},\alpha}-eV_b) - f(\xi_{\mathbf{p},\alpha}) \right],
\label{eq:final_current}
\end{equation}
where $\mathbf{p}$ runs over the extended mBZ of the sample for each valley, and the wavefunctions and dispersions of the tip are defined relative to the rotated Dirac point of the tip, denoted as $\tau\mathbf{K}_{\theta,n}$ and defined below. $\ket{u^{S,\text{top}}_{\mathbf{p},\tau,\beta}}$ denotes the sample eigenfunction, projected to the top layer, valley $\tau$, and extended mBZ momentum $\mathbf{p}$. In the case of spin degeneracy, an additional factor of 2 should be included.

Assuming the tip is rotated by angle $\theta$ relative to the sample top layer, the tip Dirac point positions under the three Umklapp processes, relative to the sample momenta, following from lattice momentum conservation are
\begin{equation}
    \mathbf{K}_{\theta,n}=\mathcal{R}(\theta)(\mathbf{K}_T+\mathbf{G}_n)-\mathbf{G}_n
    =\mathbf{K}_T+(\mathcal{R}(\theta)-\mathcal{I})(\mathbf{K}_T+\mathbf{G}_{n}),
\end{equation}
where $\mathcal{R}(\theta)$ is a rotation matrix, $\mathbf{K}_T=\tfrac{4\pi}{3a}(1,0)^T$ is the Dirac point of the sample top layer when the $K$ point is aligned with $+x$, and $a$ is the graphene lattice constant. $\mathbf{G}_1=(0,0)^T$ and $\mathbf{G}_{2,3}=\tfrac{4\pi}{3a}(-3/2,\pm\sqrt{3}/2)^T$ connect the $K$ point to the two other equivalent $K$ points. For small twist angles, $(\mathcal{R}(\theta)-\mathcal{I})(\mathbf{K}_T+\mathbf{G}_{n})$ is approximately parallel to the moiré vectors $\mathbf{q}_{n}$ of TBG, such that \cref{eq:dirac_tip}  in the main text holds.

Finally, momentum in the extended mBZ ($\mathbf{p}$) can be folded into the first mBZ ($\mathbf{k}$) using moiré reciprocal lattice vectors. Thus, $\ket{u^{S,\text{top}}_{\mathbf{p},\tau,\beta}}$ denotes a component of the multi-component Bloch wavefunction $\ket{u^{S}_{\mathbf{k},\beta}}$ obtained by diagonalizing the TBG Hamiltonian.

\section{Deriving the singularities in the QTM current derivatives}
\label{app:dirac_sing}
\subsection{Singularity from the tip Dirac point crossing normal bands}
\label{app:dirac_sing_deriv}
We review the derivation of the singularities in $d^2 I/dV_b^2$ that arise when the tip Dirac point crosses the sample bands, following Ref.~\cite{diracscan}.

We begin from the current equation, \cref{eq:tun_cur}
\begin{equation}
I=\frac{2\pi e}{\hbar}
\sum_{\mathbf{k}_T,\mathbf{k}_S}\sum_{\alpha,\beta}
\bigl|T_{\alpha\beta}(\mathbf{k}_T,\mathbf{k}_S)\bigr|^2
\int d\omega\,
A_{T,\alpha}(\mathbf{k}_T,\omega+eV_b)\,
A_{S,\beta}(\mathbf{k}_S,\omega)\,[f(\omega)-f(\omega+eV_b)].
\end{equation}
We assume infinite lifetimes and noninteracting states in both electrodes, so $A_{S,\beta}(\mathbf{k},\omega)=\delta(\omega-E_{\mathbf{k},\beta})$ and $A_{T,\alpha}(\mathbf{k},\omega)=\delta(\omega-\xi_{\mathbf{k},\alpha})$, where $\xi_{\mathbf{k},\alpha}$ and $E_{\mathbf{k},\beta}$ are the tip and sample dispersions relative to $\mu_T$ and $\mu_S$, respectively. After integrating over $\omega$, we replace the momentum sums by integrals. In addition, in-plane momentum is conserved, $\mathbf{k}\equiv\mathbf{k}_S+\mathbf{G}_S=\mathbf{k}_T+\mathbf{G}_T$, and incommensurate Umklapp channels contribute incoherently, so the tunneling matrix depends only on $\mathbf{k}$ (see discussion around~\cref{eq:tun_amp_def}). We obtain
\begin{equation}
I=\frac{2\pi e\Omega}{\hbar}
\sum_{\alpha,\beta}
\int\!\frac{d^2k}{(2\pi)^2}\,|T_{\alpha\beta}(\mathbf{k})|^2\,
A_{S,\beta}(\mathbf{k},\xi_{\mathbf{k},\alpha}-eV_b)\,
\bigl[f(\xi_{\mathbf{k},\alpha}-eV_b)-f(\xi_{\mathbf{k},\alpha})\bigr],
\end{equation}
with $\Omega$ the junction area. Inserting $A_{S,\beta}$ gives
\begin{equation}
I=\frac{2\pi e\Omega}{\hbar}
\sum_{\alpha,\beta}
\int\!\frac{d^2k}{(2\pi)^2}\,|T_{\alpha\beta}(\mathbf{k})|^2\,
\delta(\xi_{\mathbf{k},\alpha}-eV_b-E_{\mathbf{k},\beta})\,
\bigl[f(\xi_{\mathbf{k},\alpha}-eV_b)-f(\xi_{\mathbf{k},\alpha})\bigr].
\label{eq:current2}
\end{equation}
For the tip, we use the graphene linear dispersion, $\xi_{\mathbf{k},\alpha}=\alpha\hbar v_D|\mathbf{k}|-\mu_T$, and restrict the integral to a small neighborhood of the tip Dirac point (centered at $\mathbf{k}=0$) at $T=0$. This is done under the assumption that the main contribution to the differential conductance originates from this region, since the contour describing the intersection between the graphene bands and the sample changes abruptly. Then the factor $f(\xi_{\mathbf{k},\alpha}-eV_b)-f(\xi_{\mathbf{k},\alpha})$ can be replaced by $f(-eV_b-\mu_T)-f(-\mu_T)$, yielding
\begin{equation}
I=\frac{2\pi e\Omega}{\hbar}\bigl[f(-eV_b-\mu_T)-f(-\mu_T)\bigr]
\sum_{\alpha,\beta}\int\!\frac{d^2k}{(2\pi)^2}\,|T_{\alpha\beta}(\mathbf{k})|^2\,
\delta(\xi_{\mathbf{k},\alpha}-eV_b-E_{\mathbf{k},\beta}).
\end{equation}
We drop the band index $\beta$ for brevity. The singular contribution arises from a limited region near the tip Dirac point, and we linearize the sample dispersion as
$E_{\mathbf{k}}\approx E_{\kth}+\mathbf{v}_S\!\cdot\!\mathbf{k}$, denoting $\mathbf{v}_S$ as the sample group velocity, and assume that the sample wave functions vary smoothly. We use the tip Bloch state near the Dirac point,
$\ket{\psi_{\mathbf{k},\alpha}}=\tfrac{1}{\sqrt{2}}\bigl(\begin{smallmatrix}\alpha\\ e^{i\theta_{\mathbf{k}}}\end{smallmatrix}\bigr)$, where $\theta_{\mathbf{k}}$ is measured from the axis connecting the tip Dirac point to graphene $\Gamma$ (we restrict to a single valley). We approximate the sample wavefunction as constant, so the tunneling matrix takes the form
\begin{equation}
T_{\alpha}(\mathbf{k})\approx\frac{1}{\sqrt{2}}\bigl(\alpha T_A+T_B e^{-i\theta_{\mathbf{k}}}\bigr).
\end{equation}
Combining these results,
\begin{equation}
I=\frac{2\pi e\Omega}{\hbar}\bigl[f(-eV_b-\mu_T)-f(-\mu_T)\bigr]
\sum_{\alpha}\int\!\frac{d^2k}{(2\pi)^2}
\Bigl(\frac{|T_A|^2+|T_B|^2}{2}
+\alpha\frac{T_A^*T_B}{2}e^{-i\theta_{\mathbf{k}}}
+\alpha\frac{T_AT_B^*}{2}e^{i\theta_{\mathbf{k}}}\Bigr)
\delta\!\bigl(\varepsilon-\hbar\mathbf{v}_S\!\cdot\!\mathbf{k}+\alpha\hbar v_D|\mathbf{k}|\bigr),
\end{equation}
where $\varepsilon\equiv -eV_b-E_{\kth}-\mu_T$. Defining
\begin{equation}
\mathcal{I}_1(\varepsilon)=\sum_{\alpha=\pm1}\int\!\frac{d^2k}{(2\pi)^2}\,
\delta\!\bigl(\varepsilon-\hbar\mathbf{v}_S\!\cdot\!\mathbf{k}+\alpha\hbar v_D|\mathbf{k}|\bigr),\quad
\mathcal{I}_2(\varepsilon)=\sum_{\alpha=\pm1}\alpha\int\!\frac{d^2k}{(2\pi)^2}\,
e^{i\theta_{\mathbf{k}}}\,\delta\!\bigl(\varepsilon-\hbar\mathbf{v}_S\!\cdot\!\mathbf{k}+\alpha\hbar v_D|\mathbf{k}|\bigr),
\end{equation}
we evaluate $\mathcal{I}_2$ (the evaluation of $\mathcal{I}_1$ is similar). With $\mathbf{k}=(r\cos\theta_{\mathbf{k}},r\sin\theta_{\mathbf{k}})$ and
$\mathbf{v}_S=v_S(\cos\theta_S,\sin\theta_S)$, we have
$\mathbf{v}_S\!\cdot\!\mathbf{k}=v_S r\cos(\theta_{\mathbf{k}}-\theta_S)$, and after the shift
$\theta_{\mathbf{k}}\to\theta_{\mathbf{k}}+\theta_S$,
\begin{equation}
\mathcal{I}_2(\varepsilon)=e^{i\theta_S}\sum_{\alpha=\pm1}\alpha\int_0^{2\pi}\!\frac{d\theta_{\mathbf{k}}}{(2\pi)^2}\int_0^\infty\!dr\,r\,e^{i\theta_{\mathbf{k}}}
\,\delta\!\bigl(\varepsilon-\hbar v_S r\cos\theta_{\mathbf{k}}+\alpha\hbar v_D r\bigr).
\end{equation}
Evaluating the $r$-integral yields
\begin{equation}
\mathcal{I}_2(\varepsilon)=e^{i\theta_S}\sum_{\alpha=\pm1}\alpha\int_0^{2\pi}\!\frac{d\theta_{\mathbf{k}}}{(2\pi)^2}\,
\frac{r_0(\theta_{\mathbf{k}})\,\Theta[r_0(\theta_{\mathbf{k}})]}{\hbar\,|\alpha v_D-v_S\cos\theta_{\mathbf{k}}|}\,e^{i\theta_{\mathbf{k}}},
\qquad
r_0(\theta_{\mathbf{k}})=\frac{\varepsilon}{\hbar\,(v_S\cos\theta_{\mathbf{k}}-\alpha v_D)}.
\end{equation}
For $v_D>v_S$, the delta function describes the intersection between the sample and tip bands as a closed ellipse. For a fixed sign of $\varepsilon$, exactly one branch $\alpha=\pm1$ yields $r_0>0$ at each angle, so the $\Theta$–selection amounts to taking half of the $\alpha$–sum with $|r_0|$,
\begin{equation}
\mathcal{I}_2(\varepsilon)=\frac{|\varepsilon|}{2\hbar^2}\,e^{i\theta_S}\sum_{\alpha=\pm1}\alpha
\int_0^{2\pi}\!\frac{d\theta_{\mathbf{k}}}{(2\pi)^2}\,
\frac{e^{i\theta_{\mathbf{k}}}}{(v_S\cos\theta_{\mathbf{k}}-\alpha v_D)^2}.
\end{equation}
Using the following identities (for $a>b>0$)
\[
\int_0^{2\pi}\frac{e^{i\phi}\,d\phi}{(a-b\cos\phi)^2}=\frac{2\pi b}{(a^2-b^2)^{3/2}},
\qquad
\int_0^{2\pi}\frac{d\phi}{(a-b\cos\phi)^2}=\frac{2\pi a}{(a^2-b^2)^{3/2}},
\]
we obtain, for $v_D>v_S$,
\begin{equation}
\mathcal{I}_2(\varepsilon)=\frac{v_S}{v_D}\,e^{i\theta_S}\,
\frac{|\varepsilon|}{2\pi\hbar^2 v_D^2}\Bigl(1-\frac{v_S^2}{v_D^2}\Bigr)^{-3/2},
\qquad
\mathcal{I}_1(\varepsilon)=\frac{|\varepsilon|}{2\pi\hbar^2 v_D^2}\Bigl(1-\frac{v_S^2}{v_D^2}\Bigr)^{-3/2}.
\end{equation}
Collecting terms, the Dirac point contribution to the current (at zero temperature, infinite lifetime) is
\begin{equation}
I(V_b)=\bigl[f(-eV_b-\mu_T)-f(-\mu_T)\bigr]\,
\frac{\Omega e\,|T(\kth)|^2}{\hbar^3 v_D^2}\,
\Bigl(1-\frac{v_S^2}{v_D^2}\Bigr)^{-3/2}\,
\bigl|eV_b+\mu_T+E_{\kth}\bigr|,
\end{equation}
In the zero-temperature limit ($T=0$), the factor $\bigl[f(-eV_b-\mu_T)-f(-\mu_T)\bigr]$ simplifies to 
$\Theta(eV_b+\mu_T)-\Theta(\mu_T)$, where $\Theta$ denotes the Heaviside step function with $\Theta(0)=\tfrac{1}{2}$. 
The derivative of this function is constant except at $eV_b=-\mu_T$. 
Differentiating the current twice for $eV_b\neq -\mu_T$, we obtain
\begin{equation}
\frac{d^2 I}{dV_b^2}
=\bigl[f(-eV_b-\mu_T)-f(-\mu_T)\bigr]\,
\frac{2\Omega e^3 |T(\kth)|^2}{\hbar^3 v_D^2}\,
\Bigl(1-\frac{v_S^2}{v_D^2}\Bigr)^{-3/2}\,
\delta(eV_b+\mu_T+E_{\kth}),
\end{equation}
in agreement with Ref.~\cite{diracscan}. 
The tunneling matrix element is given by
\begin{equation}
|T(\kth)|^2
=\frac{1}{2}\Bigl(|T_A|^2+|T_B|^2
+\frac{v_S}{v_D}\,[T_A T_B^* e^{i\theta_S}+{\rm c.c.}]\Bigr).
\label{eq:tamp_kth}
\end{equation}

The resulting $\delta$-function singularity directly maps the sample dispersion. 
Choosing $\mu_T>0$ ($\mu_T<0$) restricts tunneling at the Dirac point to electron-like (hole-like) excitations. 
For simplicity, we set $\mu_T=0$, so that the tip Dirac point probes both particle- and hole-like excitations, 
and $-eV_b$ directly traces $E_{\kth}$. 
This choice makes the analytical expression invalid at $V_b=0$, 
where the region of invalidity is determined by the energy scale $k_BT$. 
Setting $\mu_T=0$ yields
\begin{equation}
\frac{d^2 I}{dV_b^2}
=\operatorname{sgn}(V_b)\,\frac{\Omega e^3}{\hbar^3 v_D^2}
\Bigl(1-\frac{v_S^2}{v_D^2}\Bigr)^{-3/2}
|T(\kth)|^2\,\delta(eV_b+E_{\kth}),
\end{equation}
and in the flat-band limit ($v_S\ll v_D$), this reduces to
\begin{equation}
\frac{d^2 I}{dV_b^2}
=\operatorname{sgn}(V_b)\,\frac{\Omega e^3}{\hbar^3 v_D^2}\,
|T(\kth)|^2\,\delta(eV_b+E_{\kth}).
\end{equation}
We emphasize that the choice to set $\mu_T=0$ is not mandatory; it is possible to perform two measurements at $\mu_T>0$ and $\mu_T<0$ and obtain the same information about hole- and electron-like excitations. All calculations presented in the main text are performed without these analytical approximations 
and include finite temperature and lifetime broadening effects.

\subsection{Generalization to a superconductor}
\label{app:sc_tunnel}
Superconductivity is described within the Bogoliubov–de Gennes (BdG) framework. Using the Nambu spinor $\Psi_{\mathbf{k}}^\dagger = (c_{\mathbf{k}}^\dagger, c_{-\mathbf{k}})$ (suppressing spin and valley indices), where $c_{\mathbf{k}}^\dagger$ creates an electron in the Bloch state with momentum $\mathbf{k}$, the BdG Hamiltonian reads
\begin{equation}
\mathcal{H}_{\mathrm{BdG}}
=\frac{1}{2}\sum_{\mathbf{k}}
\Psi_{\mathbf{k}}^\dagger
\begin{pmatrix}
\xi_{\mathbf{k}} & \Delta_{\mathbf{k}} \\
\Delta_{\mathbf{k}}^{\ast} & -\xi_{-\mathbf{k}}
\end{pmatrix}
\Psi_{\mathbf{k}},
\end{equation}
where $\xi_{\mathbf{k}}$ is the normal-state dispersion measured from the chemical potential and $\Delta_{\mathbf{k}}$ is the pairing potential (possibly matrix-valued)  that couples each state to its time-reversed hole counterpart. 
We focus on singlet or unitary triplet pairing, so the pairing magnitude does not depend on spin and can be represented as a scalar~\cite{mineev1999introduction,sigrist2005introduction}. For intra-band pairing with a time-reversal-symmetric normal state, diagonalization of $\mathcal{H}_{\mathrm{BdG}}$ yields $E_{\mathbf{k}}=\sqrt{\xi_{\mathbf{k}}^{2}+|\Delta_{\mathbf{k}}|^{2}}$.
The single-particle spectral function of the superconductor probed by tunneling is
\begin{equation}
A(\mathbf{k},\omega)
= |u_{\mathbf{k}}|^{2}\,\delta(\omega-E_{\mathbf{k}})
+ |v_{\mathbf{k}}|^{2}\,\delta(\omega+E_{\mathbf{k}}),
\label{eq:specfun_sc}
\end{equation} 
with the following definitions for the coherence factors: 
\begin{equation}
|u_{\mathbf{k}}|^{2}=\frac{1}{2}\!\left(1+\frac{\xi_{\mathbf{k}}}{E_{\mathbf{k}}}\right),\qquad
|v_{\mathbf{k}}|^{2}=\frac{1}{2}\!\left(1-\frac{\xi_{\mathbf{k}}}{E_{\mathbf{k}}}\right),
\label{eq:coherenc}
\end{equation}
which are the probabilities of adding and removing an electron with momentum $\mathbf{k}$, respectively. \cref{fig:QTM_b} in the main text shows a simple example of the superconducting spectral function for parabolic dispersion. We further assume a finite pair-breaking scattering rate, $\Gamma_{\mathrm{SC}}$, which broadens the spectral function by replacing the delta function with a Lorentzian of width $\Gamma_{\mathrm{SC}}$ (see discussion around~\cref{eq:spec_lorentzian}).

We follow the same steps from the previous subsection, now using the superconducting spectral function (\cref{eq:specfun_sc}). Because the coherence factors vary slowly within the narrow region controlling the singularity, we take $|u_{\mathbf{k}}|^2\simeq |u_{\kth}|^2$ and $|v_{\mathbf{k}}|^2\simeq |v_{\kth}|^2$. The Bogoliubov spectrum then replaces $E_{\mathbf{k}}$ by $\pm E_{\mathbf{k}}=\pm\sqrt{\xi_{\mathbf{k}}'^{2}+\Delta_{\mathbf{k}}^2}$, where $\xi'_\mathbf{k}$ is the normal-state dispersion of the sample. The result at $\mu_T=0$ becomes
\begin{equation}
\frac{d^2 I}{dV_b^2}
=\frac{\Omega e^3}{\hbar^3 v_D^2}\,
\Bigl(1-\frac{v_S^2}{v_D^2}\Bigr)^{-3/2}
\Bigl[-\,|T^+(\kth)|^2 |u_{\kth}|^2\,\delta(eV_b+E_{\kth})
+\,|T^-(\kth)|^2 |v_{\kth}|^2\,\delta(eV_b-E_{\kth})\Bigr],
\end{equation}
where $T^\pm(\kth)$ denote the tunneling matrix elements evaluated on the $\pm E$ branches. In the flat-band limit ($v_S\ll v_D$), $T^+(\kth)=T^-(\kth)\equiv T(\kth)$ to leading order, yielding
\begin{equation}
\frac{d^2 I}{dV_b^2}
=\frac{\Omega e^3}{\hbar^3 v_D^2}\,|T(\kth)|^2
\Bigl[-\,|u_{\kth}|^2\,\delta(eV_b+E_{\kth})
+\,|v_{\kth}|^2\,\delta(eV_b-E_{\kth})\Bigr].
\end{equation}

\subsection{Comparison of Dirac-point singularity with varying tip Fermi energy}
\label{app:change_mut}
Here, we show that the same information obtained by scanning the sample spectral function with the tip Dirac point (varying the electrostatic shift $\phi$ and plotting $d^2I/dV_b^2$ at fixed $\mu_T$ and $\mu_S$) can also be extracted from $dI/dV_b$ by sweeping $\mu_T$ while keeping $\phi$ and $\mu_S$ fixed.

Under this protocol, terms of the form $\xi_{\mathbf{k},\alpha}-eV_b=\epsilon_{\mathbf{k},\alpha}-\mu_T-eV_b$, where $\epsilon_{\mathbf{k},\alpha}$ denotes the bare dispersion, are independent of $V_b$ when $\phi$ is fixed. We set $\phi=\mu_S$, so that the tip Fermi surface at zero bias collapses to a point at the sample Fermi energy (at the middle of a superconducting gap). Differentiating \cref{eq:current2} with respect to $V_b$ yields
\begin{equation}
\left.\frac{dI}{dV_b}\right|_{\phi=\mu_S}
=-\frac{2\pi e\,\Omega}{\hbar}
\sum_{\alpha,\beta}
\int\!\frac{d^2k}{(2\pi)^2}\,|T_{\alpha\beta}(\mathbf{k})|^2\,
\delta\!\bigl(\xi_{\mathbf{k},\alpha}-eV_b-E_{\mathbf{k},\beta}\bigr)\,
\frac{\partial f(\xi_{\mathbf{k},\alpha})}{\partial V_b}.
\end{equation}
At $T=0$ and fixed $\phi$, we have
$\frac{\partial f(\xi_{\mathbf{k},\alpha})}{\partial V_b}
=-e\,\frac{\partial f}{\partial\mu_T}
=-e\,\delta(\xi_{\mathbf{k},\alpha})$.
As before, we focus on the case where the sample dispersion is relatively flat, so that the intersection between the tip and sample dispersions forms an ellipse confined to a single Dirac branch. Dropping band indices and, for a small tip Fermi circle, approximating $|T_{\alpha\beta}(\mathbf{k})|^2\simeq|\tilde T|^2$ near the tip Dirac point, we obtain
\begin{equation}
\left.\frac{dI}{dV_b}\right|_{\phi=\mu_S}
=\frac{2\pi e^2\,\Omega}{\hbar}\,|\tilde T|^2
\int\!\frac{d^2k}{(2\pi)^2}\,
\delta\!\bigl(\xi_{\mathbf{k}}-eV_b-E_{\mathbf{k}}\bigr)\,
\delta(\xi_{\mathbf{k}}).
\end{equation}
Replacing the integral over $d^2k$ with an integral over $\xi_{\mathbf{k}}$, and performing the integration, we obtain
\begin{equation}
\left.\frac{dI}{dV_b}\right|_{\phi=\mu_S}
=n_T(\mu_T)\frac{2\pi e^2\,\Omega}{\hbar}\,|\tilde T|^2\,
\delta\!\bigl(eV_b+E_{\mathbf{k}}\bigr).
\end{equation}
Here, $n_T(\mu_T)$ is the density of states in the tip at $\mu_T$. Thus, sweeping $\mu_T$ at fixed $\phi$ scans the sample dispersion as a function of $V_b$, and the signatures already appear in the first derivative of the current. Using the graphene density of states $n_T(\mu_T)=\frac{|\mu_T|}{2\pi(\hbar v_D)^2}$ gives
\begin{equation}
\left.\frac{dI}{dV_b}\right|_{\phi=\mu_S}
=|\mu_T|\frac{ e^2\,\Omega}{\hbar^3v_D^2}\,|\tilde T|^2\,
\delta\!\bigl(eV_b+E_{\mathbf{k}}\bigr).
\label{eq:dIdV_mut}
\end{equation}
The locality in momentum is preserved, provided the tip Fermi surface remains small. In typical 2D materials, the measured gap is $\lesssim 1~\si{meV}$. Increasing the tip chemical potential to $\mu_T=1~\si{meV}$ corresponds to a tip Fermi wave vector
$k_F=\mu_T/\hbar v_D\approx 1.5\times10^{-3}~\si{nm^{-1}}$, i.e., an angular averaging of only $\sim 0.005^\circ$ in terms of the tip rotation. Hence, the finite $k_F$ remains sufficiently small and the measurement is still local in momentum.

\begin{figure}[h]
\centering
\begin{tikzpicture}
      \node[inner sep=0] (img) {\includegraphics[width=0.7\linewidth]{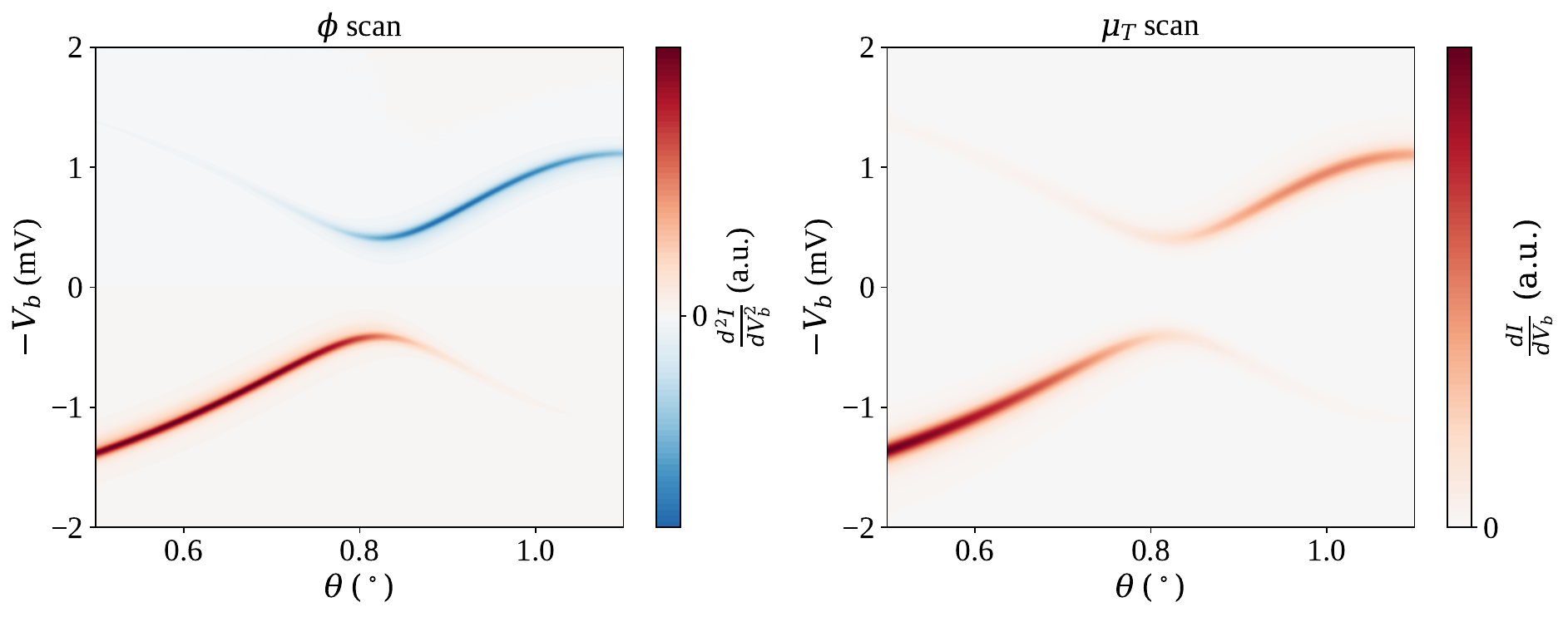}};
      \node[anchor=north west, xshift=16pt, yshift=4pt] 
        at (img.north west) {\textbf{(a)}}; 
      \node[anchor=north west, xshift=\columnwidth/3+25pt, yshift=4pt] 
        at (img.north west) {\textbf{(b)}};
    \end{tikzpicture}%
    \caption{Comparison of QTM spectra for an isotropic superconducting gap in MATBG between two measurement modes. 
    (a) $d^2I/dV_b^2$ spectrum, where $V_b$ tunes the electrostatic potential $\phi$, while $\mu_S$ and $\mu_T$ are fixed. 
    (b) $dI/dV_b$ spectrum, where $V_b$ tunes $\mu_T$, while $\mu_S$ and $\phi$ are fixed.}
    \label{fig:mut_scan}
\end{figure}

\begin{figure}[h]
\centering
\begin{tikzpicture}
      \node[inner sep=0] (img) {\includegraphics[width=0.7\linewidth]{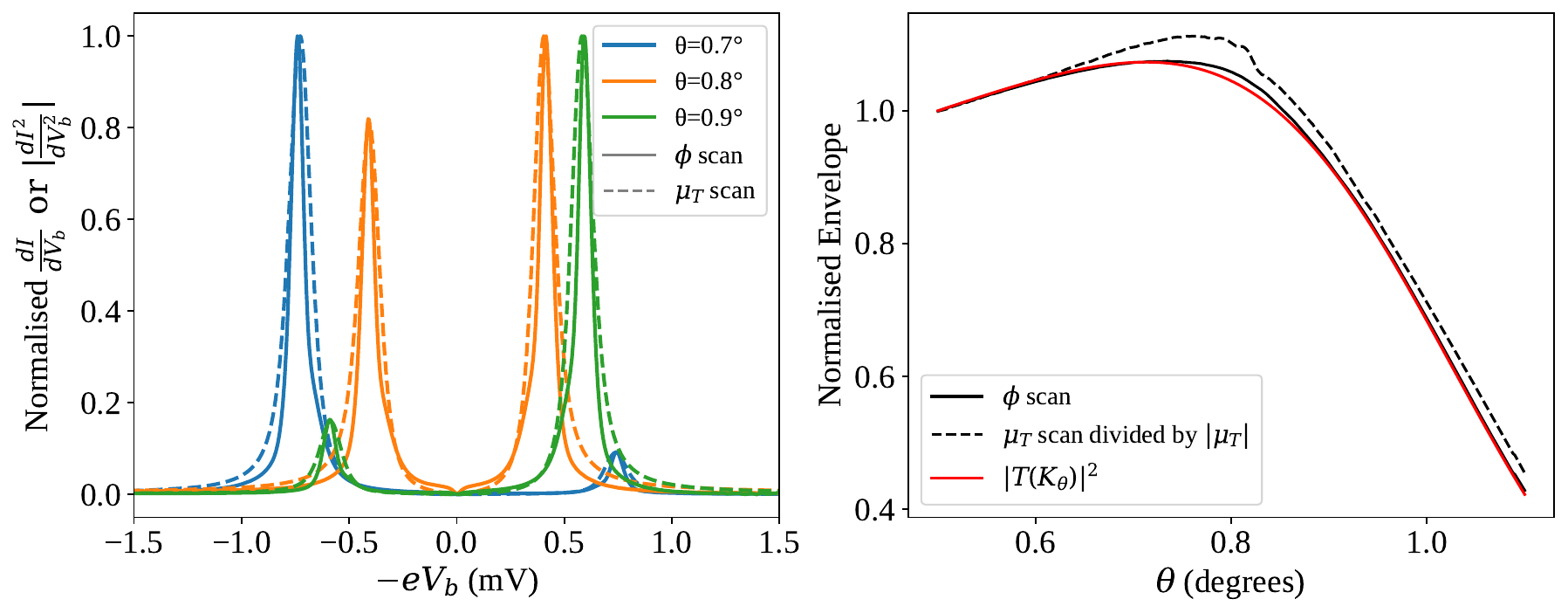}};
      \node[anchor=north west, xshift=28pt, yshift=12pt] 
        at (img.north west) {\textbf{(a)}}; 
      \node[anchor=north west, xshift=\columnwidth/3+34pt, yshift=12pt] 
        at (img.north west) {\textbf{(b)}};
    \end{tikzpicture}%
    \caption{(a) Normalized line cuts from \cref{fig:mut_scan}, comparing the peaks appearing in $dI/dV_b$ for $\mu_T$ scans and in $d^2I/dV_b^2$ for $\phi$ scans. 
    (b) The envelope of the combined peak intensities for both modes. 
    The envelope for the $\mu_T$ scan is further corrected by a factor of $|\mu_T|$ to account for the tip density of states affecting the measured conductance. The tunneling matrix element squared evaluated at the Dirac point is shown in red}
    \label{fig:mut_scan2}
\end{figure}

A numerical comparison between the two QTM spectroscopy modes is shown in \cref{fig:mut_scan,fig:mut_scan2}. 
Qualitatively, the spectral features are similar in both cases, and the bias voltage $V_b$ traces the sample spectral function in the same way. 
Quantitatively, the ratio of peak magnitude at specific twist angles matches the ratio of the coherence peaks (see main text) in both modes.
The overall signal intensity follows the tunneling matrix element, while in the $\mu_T$-scanning mode, it is additionally modulated by the density of states of the tip. 

\FloatBarrier
\subsection{Broadening effects in the QTM scans}

We review here the possible scanning variables in the QTM and the corresponding broadening mechanisms.  
The spectral-function part of the current calculation is usually written as  
$A_T(\mathbf{k},\omega+eV_b)A_S(\mathbf{k},\omega)\sim f(\omega-\phi+\mu_S-\epsilon_{T,\mathbf{k}})\cdot g(\omega-(\epsilon_{S,\mathbf{k}}-\mu_S)),$
where $f$ and $g$ are smooth functions, and $\epsilon_{S/T,\mathbf{k}}$ denote the tip and sample bare dispersions.  
This term, together with momentum conservation, dictates that the contribution to the current originates from the overlap between the sample and tip bands in both momentum and energy space.  
Therefore, the intersection between the bands—arising from this term—depends only on $\phi$.  
This consideration (together with the derivations in the previous subsections) leads to the following conclusions:

\begin{itemize}
    \item When scanning $\phi$ with a finite tip chemical potential ($\mu_T \neq 0$) far from the Dirac point, the crossing of the tip Dirac point induces a singularity in $d^{2}I/dV^{2}$.  
    In this regime, the Fermi function is relatively flat, and the temperature does not significantly broaden the feature.  
    As $\phi$ varies, the relative alignment of the tip and sample bands shifts, and the tunneling current as a function of bias voltage behaves as a convolution of their spectral functions, $A_T(\mathbf{k},\omega+eV_b)\circledast A_S(\mathbf{k},\omega)$.  
    Consequently, the feature width is determined by the sum of the tip and sample broadenings and remains insensitive to temperature.

    \item When scanning $\mu_T$ at fixed $\phi$, assuming that $\mu_T$ remains small and that the sample bands are relatively flat, a singularity in $dI/dV$ arises when the tip chemical potential crosses the intersection contour of the tip and sample bands.  
    In this case, the tunneling current is proportional to $A_T(\mathbf{k},\omega+eV_b)\cdot A_S(\mathbf{k},\omega)$ with fixed $\phi$.  
    Since the spectral functions do not shift relative to each other, no convolution broadening occurs, and the linewidth is determined mainly by the temperature through the Fermi function and the product of two Lorentzians. For large tip broadening with small sample broadening, the response becomes effectively insensitive to the tip broadening, though the finite Fermi-circle radius introduces additional momentum uncertainty.

    \item When scanning $\phi$ with $\mu_T = 0$, both broadening mechanisms contribute simultaneously, requiring low temperature and minimal tip broadening to obtain sharp features.  
    In this configuration, the interpretation is simplified because the $d^{2}I/dV^{2}$ singularity remains localized in momentum space (at the Dirac point), allowing both electron- and hole-like excitations to be probed.  
    We adopt this scanning mode for the analysis presented in the main text, although the first and second scanning schemes may be advantageous when specifically aiming to suppress temperature- or tip-induced broadening, respectively.
\end{itemize}

\section{Models for MATBG}
\label{app:tbg_models}
\subsection{Review of the Bistritzer–MacDonald model}
\label{app:BM}
We briefly review the continuum Bistritzer–MacDonald (BM) model of twisted bilayer graphene (TBG)~\cite{bistritzer2011moire}, following the notations and derivations of Refs.~\cite{song_allmagic,thfsong}. The low-energy states originate from electronic states near the graphene Dirac points of the top and bottom layers, located at valleys $K$ and $K'$. Interlayer hopping generates a moiré Brillouin zone (mBZ) onto which the eigenstates are folded, producing moiré bands. The hexagonal mBZ is described by the reciprocal lattices
\begin{equation}
    \mathbf{Q}_+ = \{ \mathbf{q}_2 + n_1 \mathbf{b}_{M1} + n_2 \mathbf{b}_{M2} \}, 
    \qquad 
    \mathbf{Q}_- = \{ -\mathbf{q}_2 + n_1 \mathbf{b}_{M1} + n_2 \mathbf{b}_{M2} \},
\end{equation}
with $n_{1,2}\in\mathbb{Z}$ and moiré reciprocal lattice vectors $\mathbf{b}_{M1}=\mathbf{q}_2-\mathbf{q}_1$ and $\mathbf{b}_{M2}=\mathbf{q}_3-\mathbf{q}_1$. The three vectors $\mathbf{q}_j$ connect the Dirac points of the two rotated layers.

We label the creation operators by $c^\dagger_{\mathbf{k},\mathbf{Q},a,\tau}$, where $\mathbf{k}$ is the crystal momentum in the mBZ, $\mathbf{Q}$ is a reciprocal vector of the triangular moiré lattice ($\mathbf{Q}\in\mathbf{Q}_+\oplus\mathbf{Q}_-$), $a$ denotes the graphene sublattice, and $\tau=\pm$ labels the valleys ($+$ for $K$, $-$ for $K'$). The spin index is suppressed. These operators relate to monolayer tight-binding operators via
\begin{align}
c^\dagger_{\mathbf{k},\mathbf{Q}\in\mathbf{Q}^\pm,a,\tau=\pm}
&= \frac{1}{\sqrt{N_{\text{tot}}}} 
\sum_{\mathbf{R}\in \text{top}}
e^{i(\tau\mathbf{K}_T+\mathbf{k}-\mathbf{Q})\cdot (\mathbf{R}+\mathbf{t}_a)} 
c^\dagger_{\mathbf{R},a}, \\[1ex]
c^\dagger_{\mathbf{k},\mathbf{Q}\in\mathbf{Q}^\pm,a,\tau=\mp}
&= \frac{1}{\sqrt{N_{\text{tot}}}} 
\sum_{\mathbf{R}'\in \text{bottom}}
e^{i(\tau\mathbf{K}_B+\mathbf{k}-\mathbf{Q})\cdot (\mathbf{R}'+\mathbf{t}'_a)} 
c^\dagger_{\mathbf{R}',a},
\end{align}
where $N_\text{tot}$ is the number of unit cells, $\mathbf{K}_T$ and $\mathbf{K}_B$ are the $K$-point of the top and bottom layers, $\mathbf{R}$ and $\mathbf{R}'$ run over the lattice sites in the top and bottom layers, and $\mathbf{t}_a$ ($\mathbf{t}'_a$) are the sublattice positions in the top (bottom) layer. Thus, for valley $\tau=+$ the top-layer states are expanded over $\mathbf{Q}\in\mathbf{Q}_+$ (and for $\tau=-$ over $\mathbf{Q}\in\mathbf{Q}_-$), with the opposite assignment for the bottom layer. 
In this basis, the BM Hamiltonian reads
\begin{equation}
\hat{H}_{\mathrm{BM}} =
\sum_{\tau=\pm}\sum_{\mathbf{k}\in \mathrm{mBZ}}
\sum_{a,a'}\sum_{\mathbf{Q},\mathbf{Q}'}
h^{(\tau)}_{\mathbf{Q}a,\mathbf{Q}'a'}(\mathbf{k})\,
c^\dagger_{\mathbf{k},\mathbf{Q},a,\tau}\,
c_{\mathbf{k},\mathbf{Q}',a',\tau},
\end{equation}
with
\begin{align}
h^{(+)}_{\mathbf{Q}a,\mathbf{Q}'a'}(\mathbf{k}) 
&= v_D\,(\mathbf{k}-\mathbf{Q})\cdot \bm{\sigma}\,\delta_{\mathbf{Q},\mathbf{Q}'}
+ \sum_{j=1}^3 [T_j]_{aa'}\,\delta_{\mathbf{Q},\mathbf{Q}'\pm \mathbf{q}_j}, \\
h^{(-)}_{\mathbf{Q}a,\mathbf{Q}'a'}(\mathbf{k}) 
&= \left(h^{(+)}_{-\mathbf{Q},a,-\mathbf{Q}',a'}(-\mathbf{k})\right)^*.
\end{align}
Here $v_D$ is the graphene Dirac velocity, $\bm{\sigma}=(\sigma_x,\sigma_y)$ acts on the sublattice indices, and the matrices $T_j$ encode interlayer tunneling,
\begin{equation}
T_j = w_{aa} \sigma_0 
+ w_{ab}\!\left(\sigma_x \cos\frac{2\pi}{3}(j-1) + \sigma_y \sin\frac{2\pi}{3}(j-1)\right),
\end{equation}
where $w_{aa}$ and $w_{ab}$ are the tunneling amplitudes for AA and AB/BA stacking, respectively. The Hamiltonian is block diagonal in valley $\tau$, with the two valleys related by time-reversal symmetry. In this work, we use $v_D=5.944~\si{eV\AA}$, $w_{aa}=66~\si{meV}$, and $w_{ab}=110~\si{meV}$.
For convenience, we also define the first-quantized basis
\begin{equation}
    \ket{\mathbf{k},\mathbf{Q},a,\tau}=c^\dagger_{\mathbf{k},\mathbf{Q},a,\tau}\ket{0}.
\end{equation}
Diagonalizing $\hat{H}_{\mathrm{BM}}$ yields the TBG Bloch eigenstates $\ket{u_{\mathbf{k},\beta}}$ with $\mathbf{k}\in\mathrm{mBZ}$, where $\beta$ is the band index. 

To compute a tunneling matrix element for a state in valley $\tau=\pm1$ with momentum $\mathbf{p}$ (in the extended mBZ), we fold it to the first mBZ using $\mathbf{p}=\mathbf{k}+\tau\mathbf{K}_T-\mathbf{Q}$ with $\mathbf{Q}\in\mathbf{Q}_\tau$. We evaluate the tunneling matrix element in the extended zone using the wavefunction component associated with the reciprocal vector $\mathbf{Q}$. Defining the following two-component projection of the wavefunction
\begin{equation}
\ket{u^{S,\mathrm{top}}_{\mathbf{p},\tau,\beta}}
\equiv
\begin{pmatrix}
\ket{\mathbf{k},\mathbf{Q},1,\tau}\braket{\mathbf{k},\mathbf{Q},1,\tau}{u_{\mathbf{k},\beta}}\\[2pt]
\ket{\mathbf{k},\mathbf{Q},2,\tau}\braket{\mathbf{k},\mathbf{Q},2,\tau}{u_{\mathbf{k},\beta}}
\end{pmatrix},
\label{eq:u_project}
\end{equation}
we then insert $\ket{u^{S,\mathrm{top}}_{\mathbf{p},\tau,\beta}}$ into \cref{eq:final_current}.

\subsection{Review of the Topological Heavy Fermion model}
\label{app:HF}
We review the heavy-fermion description of magic-angle TBG, following Ref.~\cite{thfsong}. In this picture the BM flat bands arise from hybridization between nearly localized orbitals centered at AA regions (``$f$'' orbitals) and topological conduction bands (``$c$'' bands). We introduce fermionic operators for the two electron kinds
\begin{equation}
f^\dagger_{\mathbf{R},\alpha,\tau,s}, \qquad 
c^\dagger_{\mathbf{k},\beta,\tau,s},
\end{equation}
where $\mathbf{R}$ labels the AA moiré sites, $\alpha=1,2$ indexes the two localized $f$ orbitals, $\beta=1,\dots,4$ labels the $c$-bands, $\tau=\pm$ denotes the valley ($K/K'$), and $s$ is spin (suppressed when not needed).

The noninteracting heavy-fermion Hamiltonian is
\begin{equation}
\hat{H}_{\text{HF}} = H^{(c)} + H^{(fc)},
\end{equation}
with
\begin{align}
H^{(c)} &= \sum_{|\mathbf{k}|<\Lambda_c}\sum_{\beta,\beta',\tau,s}
H^{(c,\tau)}_{\beta\beta'}(\mathbf{k}) \,
c^\dagger_{\mathbf{k},\beta,\tau,s} c_{\mathbf{k},\beta',\tau,s}, \\[2pt]
H^{(fc)} &= \frac{1}{\sqrt{N}}
\sum_{|\mathbf{k}|<\Lambda_c}\sum_{\mathbf{R}}\sum_{\alpha,\beta,\tau,s}
\Big( e^{i\mathbf{k}\cdot\mathbf{R}}\, V^{(fc,\tau)}_{\alpha\beta}(\mathbf{k}) \,
f^\dagger_{\mathbf{R},\alpha,\tau,s}\, c_{\mathbf{k},\beta,\tau,s} + \text{h.c.}\Big),
\end{align}
where $N$ is the number of moiré unit cells and $\Lambda_c$ is a momentum cutoff for the conduction bands. The $f$ electrons are assumed to have zero dispersion.

The $c$-electrons Hamiltonian is
\begin{equation}
H^{(c,\tau)}(\mathbf{k}) =
\begin{pmatrix}
0 & v_\star (\tau k_x \sigma_0 + i k_y \sigma_z) \\[2pt]
v_\star (\tau k_x\sigma_0 - i k_y\sigma_z) & M \sigma_x
\end{pmatrix},
\label{eq:Hc}
\end{equation}
where $\sigma_i$ are the Pauli matrices, $v_\star$ is an effective velocity, and $M$ is a band-splitting mass.

The $f$–$c$ hybridization is modeled as
\begin{equation}
V^{(fc,\tau)}(\mathbf{k}) = e^{-\frac{|\mathbf{k}|^2\lambda^2}{2}}
\begin{pmatrix}
\gamma\,\sigma_0 + v'_\star (\tau k_x \sigma_x + k_y \sigma_y) & 0_{2\times 2}
\end{pmatrix},
\label{eq:Hfc}
\end{equation}
where $\gamma$ and $v'_\star$ parameterize the hybridization and $\lambda$ is a fitted damping scale set by the localized $f$ orbitals (see~\cref{table:hf_params} for the adopted parameters). In Ref.~\cite{thfsong}, the $f$-electron wave functions are obtained as maximally localized Wannier functions, which span most of the flat bands except near the $\gamma$ point. Wannierization yields matrix elements $\tilde v^{(\tau)}_{\mathbf{Q}a,\alpha}(\mathbf{k})$ in a plane-wave basis $\ket{\mathbf{k},\mathbf{Q},a,\tau}$ (defined in the previous subsection). The $c$ electrons are defined as the remaining four bands after projecting onto the lowest six BM bands and removing the $f$ sector. The transformations between the $f/c$ basis and the plane-wave basis are given by
\begin{equation}
f^\dagger_{\mathbf{k}\alpha\tau}
=\frac{1}{\sqrt{N}} \sum_\mathbf{R} e^{i\mathbf{k}\cdot \mathbf{R}}f^\dagger_{\mathbf{R}\alpha\tau}
=\sum_{\mathbf{Q},a} \tilde v^{(\tau)}_{\mathbf{Q}a,\alpha}(\mathbf{k})\,c^\dagger_{\mathbf{k},\mathbf{Q},a,\tau},
\label{eq:f_trans}
\end{equation}
and
\begin{equation}
c^\dagger_{\mathbf{k}\beta\tau}
=\sum_{\mathbf{Q},a} \tilde u^{(\tau)}_{\mathbf{Q}a,\beta}(\mathbf{k})\, c^\dagger_{\mathbf{k},\mathbf{Q},a,\tau}.
\label{eq:c_trans}
\end{equation}
Note that the $c$-electrons creation operator, $c^\dagger_{\mathbf{k}\beta\tau}$, is not periodic in the reciprocal space, whereas $f^\dagger_{\mathbf{k}\alpha\tau}$ is periodic because it is defined on the lattice. To compute tunneling matrix elements, we use the analytical approximations and fitted parameters of Ref.~\cite{cualuguaru2023twisted}. For the $f$-electrons,
\begin{align}
\begin{split}
\tilde v^{(\tau)}_{\mathbf{Q}1,1}(\mathbf{k}) 
&= \alpha_1 \sqrt{\frac{2\pi \lambda_1^2}{\Omega_M \mathcal{N}_{f,\mathbf{k}}}}
\, e^{\,i\frac{\pi}{4}\zeta_{\mathbf{Q}} - \tfrac{1}{2}(\mathbf{k}-\mathbf{Q})^2 \lambda_1^2}, \\[1ex]
\tilde v^{(\tau)}_{\mathbf{Q}2,1}(\mathbf{k}) 
&= \alpha_2 \sqrt{\frac{2\pi \lambda_2^4}{\Omega_M \mathcal{N}_{f,\mathbf{k}}}}
\,\zeta_{\mathbf{Q}}\,[\,i\tau(k_x-Q_x) - (k_y-Q_y)\,]\,
e^{\,i\frac{\pi}{4}\zeta_{\mathbf{Q}} - \tfrac{1}{2}(\mathbf{k}-\mathbf{Q})^2 \lambda_2^2}, \\[1ex]
\tilde v^{(\tau)}_{\mathbf{Q}1,2}(\mathbf{k}) 
&= \alpha_2 \sqrt{\frac{2\pi \lambda_2^4}{\Omega_M \mathcal{N}_{f,\mathbf{k}}}}
\,\zeta_{\mathbf{Q}}\,[\,-i\tau(k_x-Q_x) - (k_y-Q_y)\,]\,
e^{-\,i\frac{\pi}{4}\zeta_{\mathbf{Q}} - \tfrac{1}{2}(\mathbf{k}-\mathbf{Q})^2 \lambda_2^2}, \\[1ex]
\tilde v^{(\tau)}_{\mathbf{Q}2,2}(\mathbf{k}) 
&= \alpha_1 \sqrt{\frac{2\pi \lambda_1^2}{\Omega_M \mathcal{N}_{f,\mathbf{k}}}}
\, e^{-\,i\frac{\pi}{4}\zeta_{\mathbf{Q}} - \tfrac{1}{2}(\mathbf{k}-\mathbf{Q})^2 \lambda_1^2},
\end{split}
\end{align}
with normalization
\begin{equation}
\mathcal{N}_{f,\mathbf{k}} = 
\alpha_1^2 \frac{2\pi \lambda_1^2}{\Omega_M}
\sum_{\mathbf{Q}} e^{-(\mathbf{k}-\mathbf{Q})^2 \lambda_1^2}
+ \alpha_2^2 \frac{2\pi \lambda_2^4}{\Omega_M}
\sum_{\mathbf{Q}} (\mathbf{k}-\mathbf{Q})^2 e^{-(\mathbf{k}-\mathbf{Q})^2 \lambda_2^2}.
\end{equation}
Here $\zeta_{\mathbf{Q}}=\pm1$ for $\mathbf{Q}\in\mathbf{Q}_\pm$, $\Omega_M$ is the moiré unit-cell area, and the parameters for the wavefunctions are given in~\cref{table:hf_params}.

\begin{table}[h!]
\centering
\begin{tabular}{cc||ccccc|cccc}
\hline
${w_{aa}}/{w_{ab}}$ & $\theta~(^\circ)$ 
& $\gamma~(\si{meV})$ & $v'_\star~(\si{eV\AA})$ & $v_\star~(\si{eV\AA})$ & $M~(\si{meV})$ & $\lambda~(a_M)$ 
& $\lambda_1~(a_M)$ & $\lambda_2~(a_M)$ & $\alpha_1$ & $\alpha_2$ \\
\hline
\hline
0.8 & 1.05 
& $-24.75$ & $1.623$ & $-4.303$ & $3.697$ & $0.3375$ 
& $0.1791$ & $0.1910$ & $0.8193$ & $0.5734$ \\
\hline
0.6 & 1.10 
& $-60.527$ & $1.604$ & $-4.753$ & $-3.526$ & $0.376$ 
& $0.215$ & $0.209$ & $0.892$ & $0.452$ \\
\hline
\end{tabular}
\caption{Model parameters for different values of $w_{aa}/w_{ab}$ and twist angle $\theta_{\mathrm{TBG}}$. $a_M$ is the moiré lattice constant. The parameters are adopted from~\cite{thfsong,cualuguaru2023twisted}.}
\label{table:hf_params}
\end{table}

For the $c$-electrons we use the analytic form at $\mathbf{k}=0$ (adopted from~\cite{cualuguaru2023twisted}):
\begin{align}
\begin{alignedat}{2}
\tilde u^{(\tau)}_{\mathbf{Q}1,1}(0) &= -\alpha_{c1} 
\sqrt{\frac{2\pi \lambda_{c1}^2}{\Omega_M \mathcal{N}_{c1}}}\,
e^{-i \tfrac{\pi}{4} \zeta_{\mathbf{Q}} - \tfrac{1}{2} \mathbf{Q}^2 \lambda_{c1}^2}, &
\qquad
\tilde u^{(\tau)}_{\mathbf{Q}1,2}(0) &= \alpha_{c2} 
\sqrt{\frac{\pi \lambda_{c2}^6}{\Omega_M \mathcal{N}_{c2}}}\,
(-i\tau Q_x + Q_y)^2 e^{i \tfrac{\pi}{4} \zeta_{\mathbf{Q}} - \tfrac{1}{2} \mathbf{Q}^2 \lambda_{c2}^2}, \\[0.7ex]
\tilde u^{(\tau)}_{\mathbf{Q}1,3}(0) &= \alpha_{c3} 
\sqrt{\frac{2\pi \lambda_{c3}^4}{\Omega_M \mathcal{N}_{c3}}}\,
\zeta_{\mathbf{Q}} (-i\tau Q_x + Q_y)\,
e^{-i \tfrac{\pi}{4} \zeta_{\mathbf{Q}} - \tfrac{1}{2} \mathbf{Q}^2 \lambda_{c3}^2}, &
\qquad
\tilde u^{(\tau)}_{\mathbf{Q}1,4}(0) &= \alpha_{c4} 
\sqrt{\frac{\pi \lambda_{c4}^6}{\Omega_M \mathcal{N}_{c4}}}\,
(i\tau Q_x + Q_y)^2 e^{i \tfrac{\pi}{4} \zeta_{\mathbf{Q}} - \tfrac{1}{2} \mathbf{Q}^2 \lambda_{c4}^2}, \\[0.7ex]
\tilde u^{(\tau)}_{\mathbf{Q}2,1}(0) &= \alpha_{c2} 
\sqrt{\frac{\pi \lambda_{c2}^6}{\Omega_M \mathcal{N}_{c1}}}\,
(i\tau Q_x + Q_y)^2 e^{-i \tfrac{\pi}{4} \zeta_{\mathbf{Q}} - \tfrac{1}{2} \mathbf{Q}^2 \lambda_{c2}^2}, &
\qquad
\tilde u^{(\tau)}_{\mathbf{Q}2,2}(0) &= -\alpha_{c1} 
\sqrt{\frac{2\pi \lambda_{c1}^2}{\Omega_M \mathcal{N}_{c2}}}\,
e^{i \tfrac{\pi}{4} \zeta_{\mathbf{Q}} - \tfrac{1}{2} \mathbf{Q}^2 \lambda_{c1}^2}, \\[0.7ex]
\tilde u^{(\tau)}_{\mathbf{Q}2,3}(0) &= \alpha_{c4} 
\sqrt{\frac{\pi \lambda_{c4}^6}{\Omega_M \mathcal{N}_{c3}}}\,
(-i\tau Q_x + Q_y)^2 e^{-i \tfrac{\pi}{4} \zeta_{\mathbf{Q}} - \tfrac{1}{2} \mathbf{Q}^2 \lambda_{c4}^2}, &
\qquad
\tilde u^{(\tau)}_{\mathbf{Q}2,4}(0) &= \alpha_{c3} 
\sqrt{\frac{2\pi \lambda_{c3}^4}{\Omega_M \mathcal{N}_{c4}}}\,
\zeta_{\mathbf{Q}} (i\tau Q_x + Q_y)\,
e^{i \tfrac{\pi}{4} \zeta_{\mathbf{Q}} - \tfrac{1}{2} \mathbf{Q}^2 \lambda_{c3}^2}.
\end{alignedat}
\end{align}
For $w_{aa}/w_{ab}=0.8$ and $\theta_{\text{TBG}} = 1.05^\circ$, fitting to the continuum BM solutions gives
\begin{align}
\begin{split}
\lambda_{c1}&= 0.2194 a_M, \quad \lambda_{c2} = 0.3299 a_M,\quad
\lambda_{c3} = 0.2430 a_M, \quad \lambda_{c4} = 0.2241 a_M,\\
\alpha_{c1} &= 0.3958, \quad \alpha_{c2} = 0.9183,\quad
\alpha_{c3} = 0.9257, \quad \alpha_{c4} = 0.3783,\\
\mathcal{N}_{c1}& = \mathcal{N}_{c2} = 1.2905,\quad
\mathcal{N}_{c3} = \mathcal{N}_{c4} = 1.1102,
\end{split}
\end{align}
and we find that these values also fit the BM solution for $w_{aa}/w_{ab}=0.6$, $\theta_{\text{TBG}}=1.1^\circ$. Following Ref.~\cite{cualuguaru2023twisted}, we further approximate
\begin{equation}
    \tilde u ^{(\tau)}_{\mathbf{Q}a,\beta}(\mathbf{k})\approx \tilde u^{(\tau)}_{\mathbf{Q}-\mathbf{k},a,\beta}(0).
\end{equation}
To obtain the eigenstates of the heavy-fermion model in the plane-wave basis, we first diagonalize the Hamiltonian in the $f$–$c$ basis and then transform to the plane-wave using \cref{eq:f_trans,eq:c_trans}. This yields eigenstates $\ket{u_{\mathbf{k},\beta}}$, which can be projected to the top layer component to compute tunneling matrix elements, as in the previous subsection, using~\cref{eq:u_project}.

Finally, interactions can be incorporated within a Hartree–Fock treatment. Ref.~\cite{thfsong} defines parent states with occupied $f$ electrons that define specific ground-state symmetry. For example, a valley-polarized parent state is
\begin{equation}
\ket{\mathrm{VP}^{\nu=0}_0} 
= \prod_{\mathbf{R}}
f^\dagger_{\mathbf{R}1,\tau=+}\,
f^\dagger_{\mathbf{R}2,\tau=+}\,
\ket{\mathrm{FS}},
\end{equation}
where the Fermi sea $\ket{\mathrm{FS}}$ has all lower $c$ bands filled and the $f$ and upper $c$ bands empty. Ref.~\cite{thfsong} shows that the one-shot mean-field Hamiltonian calculated using the parent state provides a good approximation to the fully self-consistent solution, and we adopt this approach. Restoring spin, we define the density matrix of the $f$-sector,
\begin{equation}
O^{f}_{\alpha \tau s, \alpha' \tau' s'} 
= \bra{\Psi} f^\dagger_{\mathbf{R}\alpha\tau s} f_{\mathbf{R}\alpha' \tau' s'} \ket{\Psi} 
= \frac{1}{N} \sum_{\mathbf{k}} 
\bra{\Psi} f^\dagger_{\mathbf{k}\alpha\tau s} f_{\mathbf{k}\alpha' \tau' s'} \ket{\Psi},
\qquad \nu_f=\mathrm{Tr}(O^f)-4,
\label{eq:f_matrix}
\end{equation}
with $\ket{\Psi}$ the chosen parent state. The interaction-induced correction in the one-shot approximation depends only on $O^f$~\cite{thfsong}. The contribution to the energies of the $f$ states is
\begin{equation}
[\overline{H}_U]_{\alpha \tau s, \alpha' \tau' s'}
= 
\Big( U_1(\nu_f + 0.5) + 6 U_2 \nu_f \Big) \delta_{\alpha,\alpha'}\delta_{\tau,\tau'}\delta_{s,s'}
- U_1 
O^f_{\alpha \tau s, \alpha \tau' s'}.
\end{equation}
The correction to $H^{(c)}$ separates into the $\beta=1,2$ conduction bands,
\begin{equation}
[\overline{H}_{W_1}]_{\beta \tau s, \beta' \tau' s'} = \nu_f W_1\delta_{\beta,\beta'}\delta_{\tau,\tau'}\delta_{s,s'},
\end{equation}
and the $\beta=3,4$ bands,
\begin{equation}
[\overline{H}_{W_3}+\overline{H}_{J}]_{\beta \tau s, \beta' \tau' s'}
= \nu_f W_3\,\delta_{\beta\beta'}\delta_{\tau\tau'}\delta_{ss'}
- J \delta_{\beta \beta'}\delta_{\tau\tau'}\!\left(
O^f_{\beta-2,\tau,s;\,\beta-2,\tau,s'} - \tfrac{1}{2}\delta_{ss'}\right)
+ J \delta_{\beta,\beta'}\delta_{-\tau,\tau'} O^f_{\beta-2,-\tau,s';\,\beta-2,\tau,s}.
\end{equation}
The interaction parameters are given in~\cref{table:hf_params_int}.
\begin{table}[h!]
\centering
\begin{tabular}{cc|ccccc}
\hline
${w_{aa}}/{w_{ab}}$ & $\theta~(^\circ)$ 
& $U_1~(\si{meV})$ & $J~(\si{meV})$ & $W_1~(\si{meV})$ & $W_3~(\si{meV})$ & $U_2~(\si{meV})$ \\
\hline
\hline
$0.8$ & $1.05$ 
& $57.95$ & $16.38$ & $44.03$ & $50.20$ & $2.329$ \\
\hline
$0.6$ & $1.10$ 
& $51.18$ & $20.68$ & $48.49$ & $52.84$ & $2.76$ \\
\hline
\end{tabular}
\caption{Interaction parameters for different values of $w_{aa}/w_{ab}$ and twist angle $\theta_{\mathrm{TBG}}$. Adopted from Refs.~\cite{thfsong,cualuguaru2023twisted}}
\label{table:hf_params_int}
\end{table}

\subsection{Comparing between BM and Heavy Fermion for the QTM}
\begin{figure}[h]
    \centering
    \includegraphics[width=0.7\linewidth]{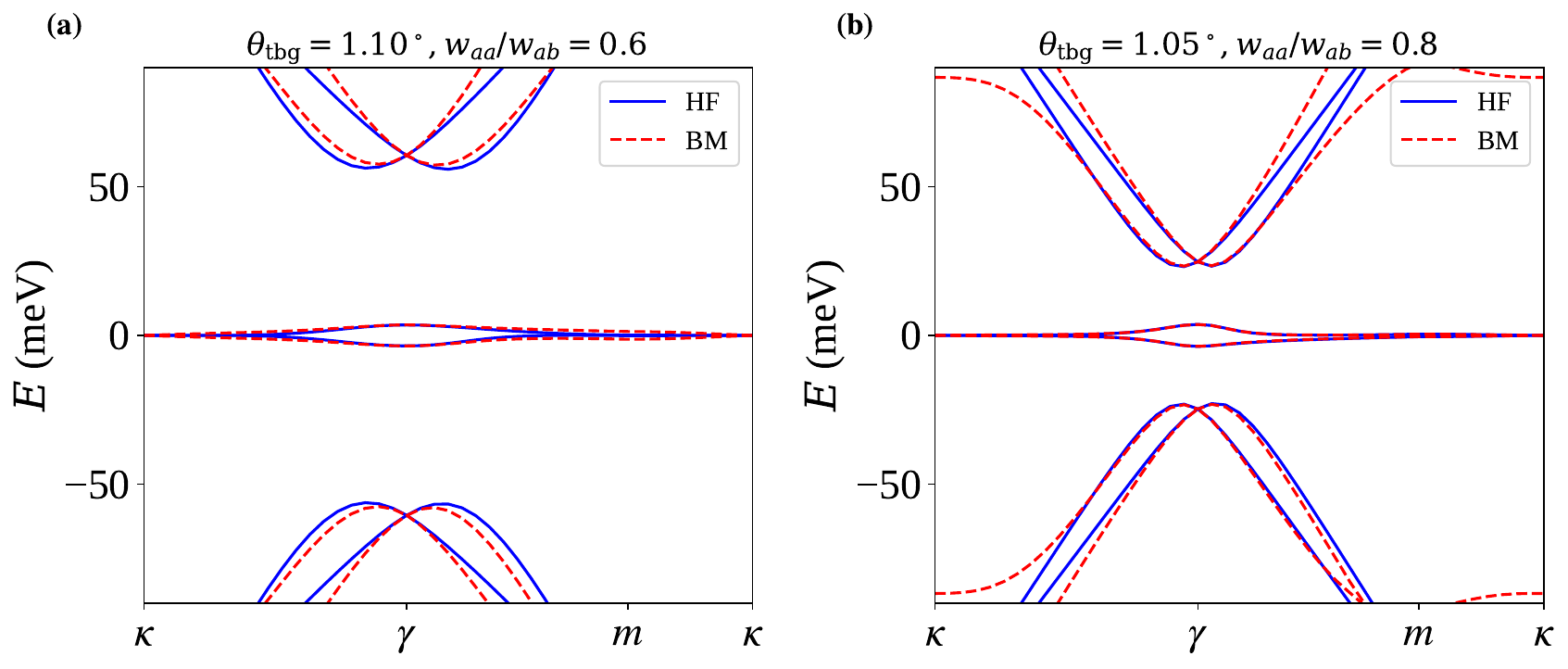}
    \caption{Comparison between the band structure of BM model and HF model along a trajectory in the mBZ. The bands presented are for a single valley. (a) for $w_{aa}/w_{ab}=0.6$ and $\theta_{\text{TBG}}=1.1^\circ$. (b) for $w_{aa}/w_{ab}=0.8$ and $\theta_{\text{TBG}}=1.05^\circ$.}
    \label{fig:hfbm_bands}
\end{figure}

\begin{figure}[h]
    \centering
    \includegraphics[width=0.7\linewidth]{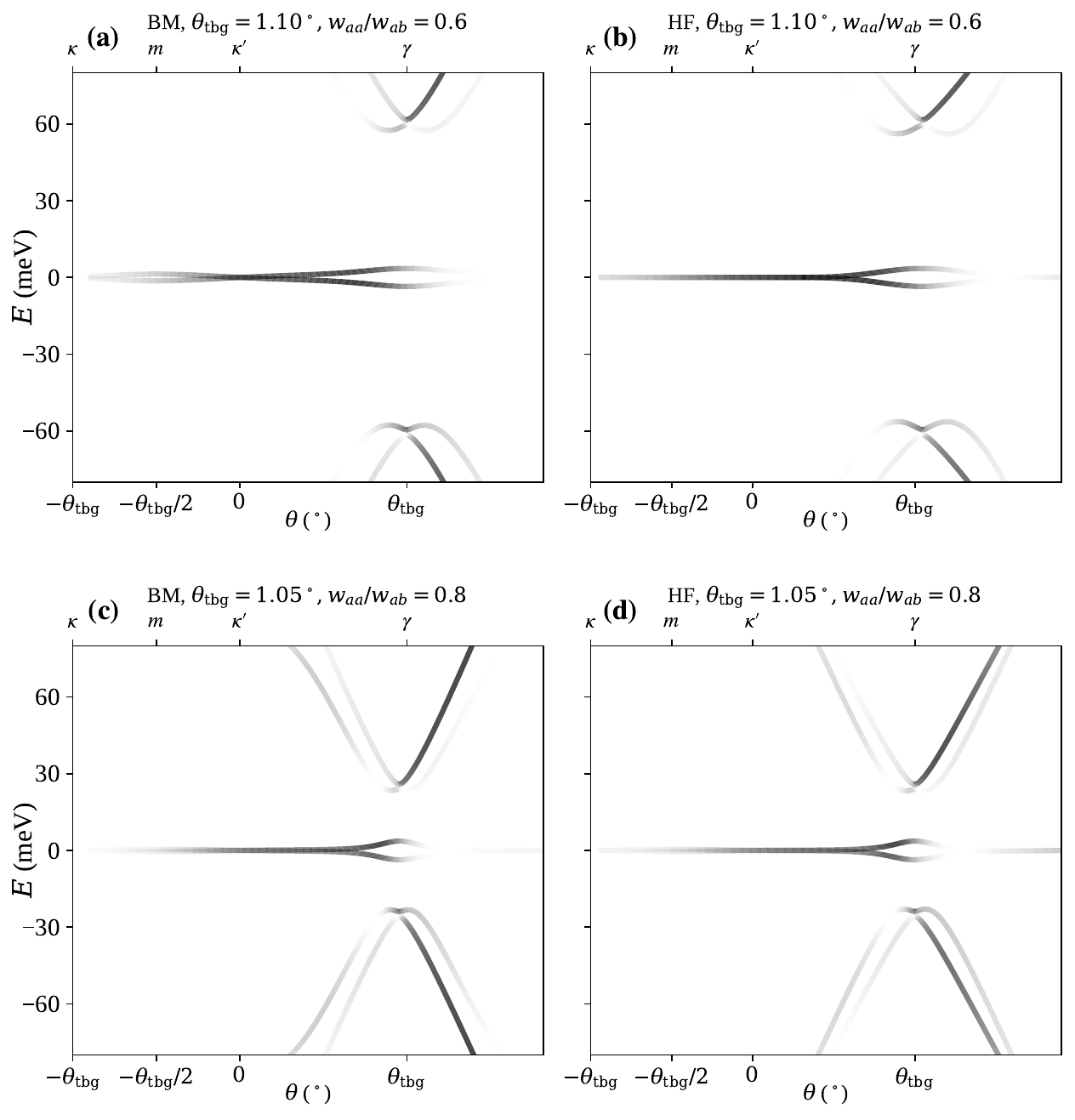}
    \caption{Comparison between the band structure and tunneling matrix element of BM model and HF model along the QTM line scan. The color intensity indicates the tunneling amplitude magnitude, where the normalization of the color intensity is the same for all figures. (a) BM model, $w_{aa}/w_{ab}=0.6,\ \theta_{\text{TBG}}=1.1^\circ$. (b) HF model, $w_{aa}/w_{ab}=0.6,\ \theta_{\text{TBG}}=1.1^\circ$. (c) BM model, $w_{aa}/w_{ab}=0.8,\ \theta_{\text{TBG}}=1.05^\circ$. (d) HF model, $w_{aa}/w_{ab}=0.8,\ \theta_{\text{TBG}}=1.05^\circ$.}
    \label{fig:hfbm_tamp}
\end{figure}

To compare the BM and HF models, we calculate the band structure (\cref{fig:hfbm_bands}) and the tunneling matrix amplitude along the QTM line scan (\cref{fig:hfbm_tamp}). We perform the comparison using the parameters from the main text, $w_{aa}/w_{ab} = 0.6$ and $\theta_{\text{TBG}} = 1.1^{\circ}$, as well as the original parameters of the HF model, $w_{aa}/w_{ab} = 0.8$ and $\theta_{\text{TBG}} = 1.05^{\circ}$~\cite{thfsong}. Overall, we find good agreement between the two models in both the eigenstates energy and the tunneling matrix amplitudes, validating the use of the HF basis as a consistent framework for calculating tunneling matrix amplitudes.

\FloatBarrier

\subsection{$\nu=-2$ Interacting Hamiltonian for MATBG}
\label{app:num2}
\begin{figure}[h]
    \centering
\includegraphics[width=0.65\linewidth]{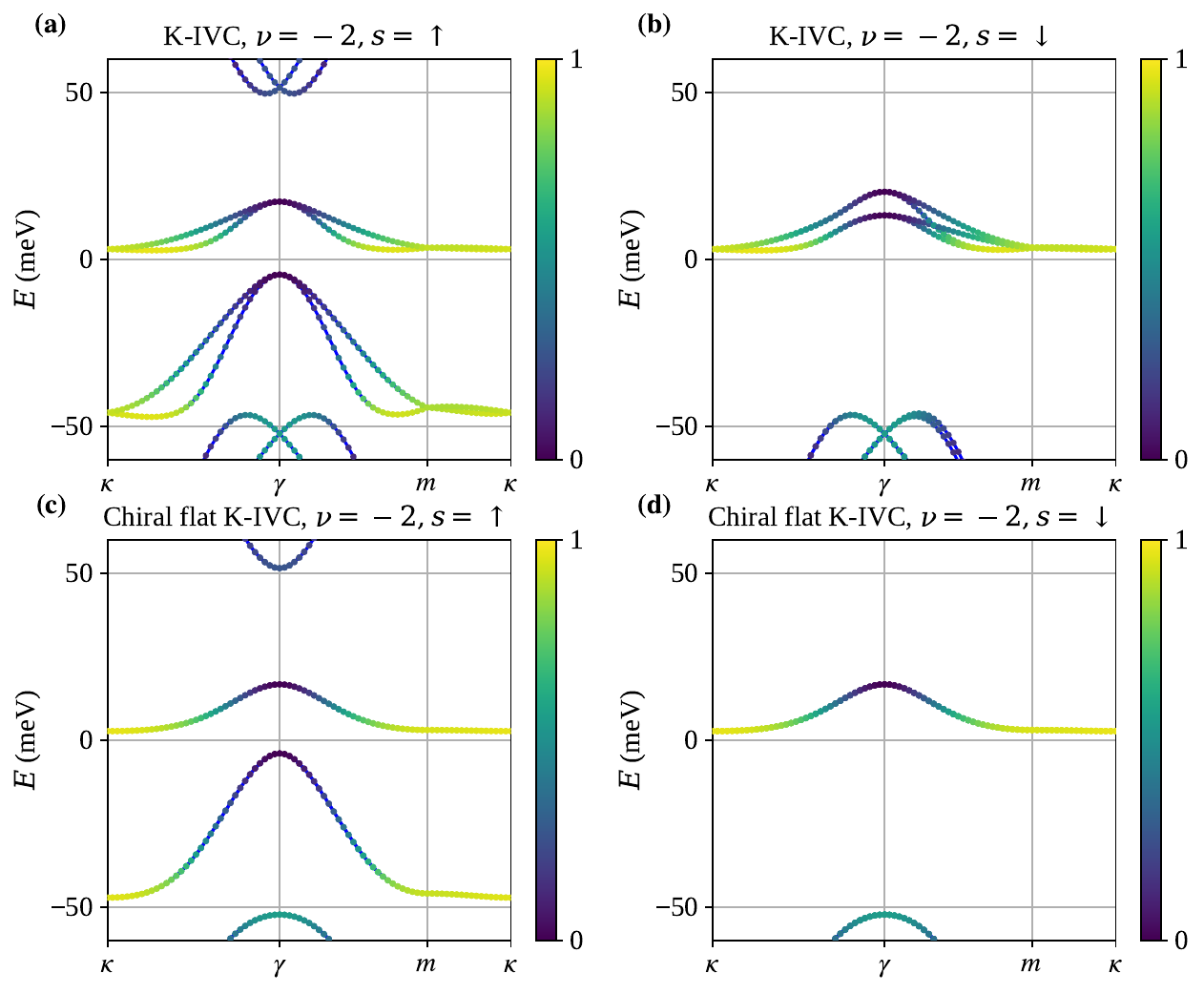}
    \caption{$\nu=-2$, K-IVC bands, for the parameters $w_{aa}/w_{ab}=0.6,\  \theta_{\text{TBG}}=1.1^\circ$. The color indicates the weight of the $f$ electrons in the eigenstates. (a-b) spin up and spin down. (c-d) spin up and spin down with the approximation $M=0,\ v_\star'=0$.}
    \label{fig:hf_kivc_bands_06}
\end{figure}

\begin{figure}[h]
    \centering
    \includegraphics[width=0.65\linewidth]{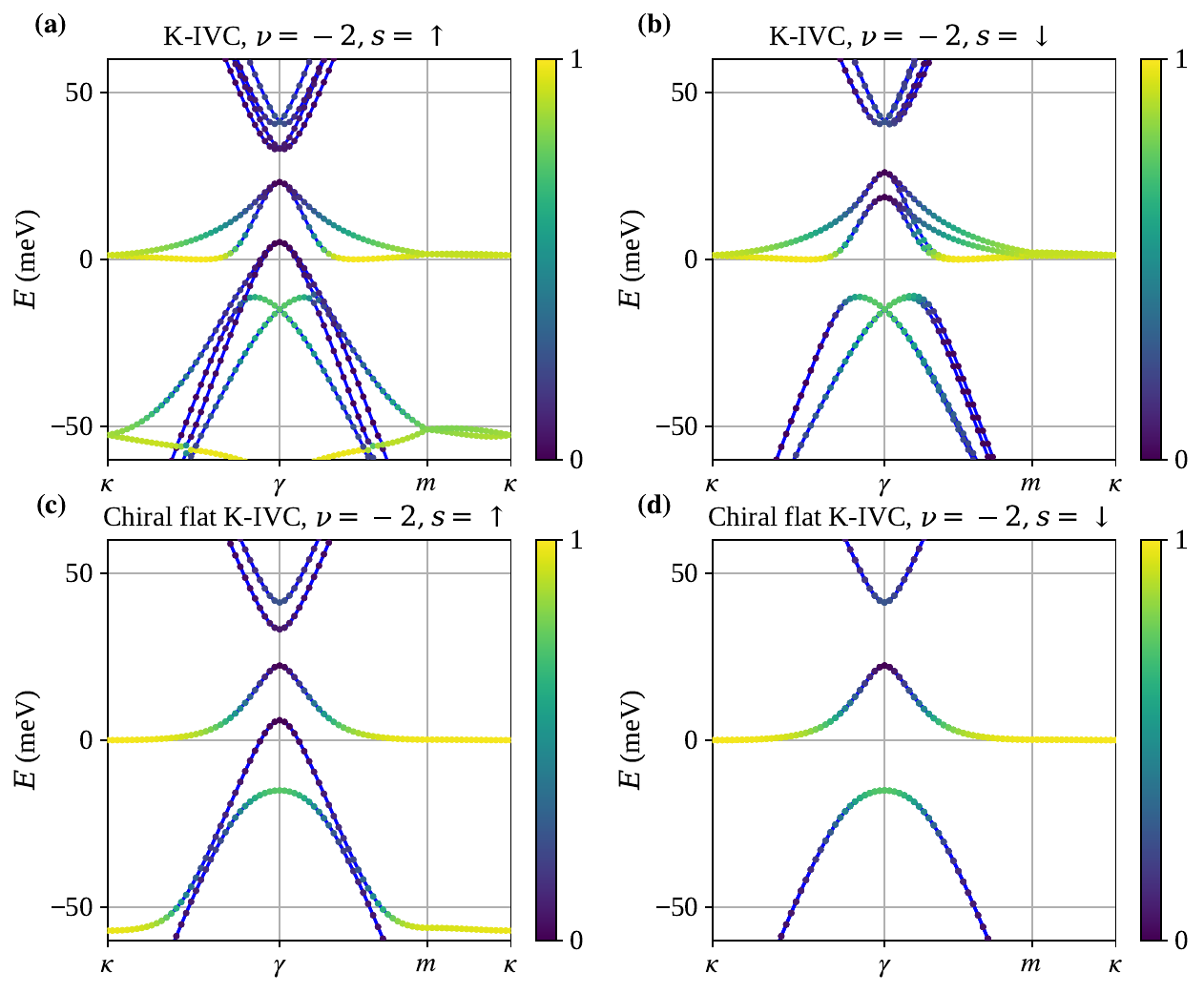}
    \caption{$\nu=-2$, K-IVC bands, for the parameters $w_{aa}/w_{ab}=0.8,\  \theta_{\text{TBG}}=1.05^\circ$. The color indicates the weight of the $f$ electrons in the eigenstates. (a-b) Spin up and spin down. (c-d) Spin up and spin down with the approximation $M=0,\ v_\star'=0$.}
    \label{fig:hf_kivc_bands_08}
\end{figure}

In the main text, we use $\nu=-2$ with K-IVC ground state as the normal state, on which superconductivity emerges. To do so, we adopt the following parent state based on the $f$ electrons~\cite{thfsong}:
\begin{equation}
    \ket{\mathrm{K\text{-}IVC}_{0}^{\nu=-2}}
= \prod_{\mathbf R}\frac{1}{2}
\bigl(f^{\dagger}_{\mathbf R,1+,\uparrow}+f^{\dagger}_{\mathbf R,2-,\uparrow}\bigr)
\bigl(-f^{\dagger}_{\mathbf R,1-,\uparrow}+f^{\dagger}_{\mathbf R,2+,\uparrow}\bigr)
\ket{\mathrm{FS}}\, 
\end{equation}
Following the definition in \cref{eq:f_matrix}, the density matrix for this parent state is spin block-diagonal and is given by
\begin{equation}
O^{(f)}_{\alpha\eta s,\;\alpha'\eta' s'} 
= \delta_{s\uparrow}\,\delta_{ss'}\!\left[\tfrac12\,\delta_{\alpha\alpha'}\delta_{\tau\tau'}
-\tfrac12\,(\sigma_y)_{\alpha\alpha'}(\tau_y)_{\tau\tau'}\right].
\end{equation}
Here, $\sigma_y,\tau_y$ are Pauli matrices acting in the $f$ and valley spaces, respectively.
In \cref{fig:hf_kivc_bands_06,fig:hf_kivc_bands_08} we present the bands for this parent state based on the one-shot approximation. To simplify the discussion, we further assume the flat and chiral limits, setting $M=v_\star'=0$, thus restoring $U(4)\times U(4)$ symmetry in the noninteracting Hamiltonian~\cite{thfsong}. This symmetry causes degeneracy in the bands. The lowest bands consist of twofold-degenerate valence bands and sixfold-degenerate conduction bands. In both cases, the valence band is spin polarized. The Hamiltonian obeys a modified spinless Kramers time-reversal symmetry~\cite{bultinck2020ground}.
\FloatBarrier

\section{Superconductor spectral function with lifetime broadening}
\label{app:lifetime}
The Green’s function of a superconductor can be written compactly using the Nambu–Gor’kov formalism~\cite{mineev1999introduction}:  
\begin{equation}
\hat G_0(\mathbf{k},\omega_n)
=-\frac{i\omega_n+\xi_{\mathbf{k}}\,\tau_{3}+\Delta_{\mathbf{k}}\tau_1}
{\omega_n^{2}+\xi_{\mathbf{k}}^{2}+\Delta_{\mathbf{k}}^{2}},
\end{equation}
where $\tau_i$ are Pauli matrices in Nambu space, and we assume intra-band pairing.  The retarded and advanced Green's functions are given by
\begin{equation}
\hat G^{R(A)}(\mathbf{k},\omega)=\hat G(\mathbf{k},\omega)|_{i\omega_n\!\to\!\omega\pm i0}.
\end{equation}
This yields
\begin{equation}
\hat G^R(\mathbf{k},\omega)
=\frac{\omega\tau_0+\xi_{\mathbf{k}}\tau_3+\Delta_{\mathbf{k}}\,\tau_1}
{\omega^{2}-\xi_{\mathbf{k}}^{2}-\Delta_{\mathbf{k}}^{2}}.
\end{equation}
Several phenomenological approaches have been developed to describe lifetime broadening in tunneling experiments on superconductors~\cite{dynes,dynestwo,dynesexp1,dynesexp2}. Here we adopt the model of Ref.~\cite{dynestwo}, which incorporates pair-breaking and pair-conserving scattering processes with rates $\Gamma$ and $\Gamma_s$, respectively. The retarded Green’s function then reads  
\begin{equation}
\hat{G}^{R}(\mathbf{k},\omega) =
\frac{\left(1 + i\Gamma_s / \Omega_{\mathbf{k}}(\omega) \right)
\left[ (\omega + i\Gamma)\tau_0 + \Delta_{\mathbf{k}}\tau_1 \right]
+ \xi_{\mathbf{k}}\tau_3}
{(\Omega_{\mathbf{k}}(\omega) + i\Gamma_s)^2 - \xi_{\mathbf{k}}^2},
\end{equation}
where
\begin{equation}
\Omega_{\mathbf{k}}(\omega) = \sqrt{(\omega + i\Gamma)^2 - \Delta_{\mathbf{k}}^2}.
\end{equation}
The normal component of the spectral function is
\begin{equation}
A(\mathbf{k},\omega)=-\frac{1}{\pi}\,\Im\, G^{R}_{11}(\mathbf{k},\omega).
\end{equation}
For simplicity, we neglect the pair-conserving rate $\Gamma_s$. The Green’s function then reduces to  
\begin{equation}
\hat{G}^{R}(\mathbf{k},\omega) =
\frac{(\omega + i\Gamma)\tau_0 + \Delta_{\mathbf{k}}\tau_1 + \xi_{\mathbf{k}}\tau_3}
{(\omega+i\Gamma)^2-\Delta_\mathbf{k}^2 - \xi_{\mathbf{k}}^2}.
\end{equation}
The resulting spectral function is
\begin{equation}
A(\mathbf{k},\omega)
= -\frac{1}{\pi}\,\Im\, G^{R}_{11}(\mathbf{k},\omega)
= \frac{1}{\pi}\left[
|u_{\mathbf{k}}|^{2}\,\frac{\Gamma}{(\omega - E_{\mathbf{k}})^{2} + \Gamma^{2}}
+ |v_{\mathbf{k}}|^{2}\,\frac{\Gamma}{(\omega + E_{\mathbf{k}})^{2} + \Gamma^{2}}
\right],
\label{eq:spec_lorentzian}
\end{equation}
with $E_\mathbf{k}=\sqrt{\xi_\mathbf{k}^2+\Delta_\mathbf{k}^2}$ and the coherence factors defined in \cref{eq:coherenc}. Thus, a finite pair-breaking rate replaces the delta-function singularities by Lorentzians centered at the quasiparticle poles.  

To recover the familiar Dynes formula for the density of states~\cite{dynes}, we assume isotropic pairing. The density of states is
\begin{equation}
    N(\omega)=\sum_\mathbf{k} -\frac{1}{\pi}\Im G_{11}^R(\mathbf{k},\omega)
    =-\frac{N_0}{\pi} \Im \int^\infty_{-\infty} d\xi_{\mathbf{k}}
    \frac{(\omega + i\Gamma)+\xi_{\mathbf{k}}}
{(\omega+i\Gamma)^2-\Delta^2 - \xi_{\mathbf{k}}^2},
\end{equation}
where $N_0$ is the normal-state density of states. The linear term in $\xi_\mathbf{k}$ vanishes after integration, leaving only the first term, which can be evaluated by contour integration. The result is the Dynes formula:  
\begin{equation}
    N(\omega)=N_0\Re\left[\frac{\omega+i\Gamma}{\sqrt{(\omega+i\Gamma)^2-\Delta^2}}\right].
\end{equation}
\section{Numerical details}
\label{app:numeric}
We use \cref{eq:final_current} in our numerical calculations. The two-dimensional momentum space is discretized to perform the summation. In the calculations, singularities in the current derivatives arise from intersections of the tip and sample bands or the Fermi surfaces with the bands. The sample bands and wavefunctions vary smoothly in $\mathbf{k}$.
Accordingly, we define a relatively coarse grid on which the eigenstates and tunneling amplitudes are computed. Linear interpolations are then constructed from this coarse grid. Finally, the current in \cref{eq:final_current} is evaluated on a finer momentum grid as a function of $V_b$, using interpolation together with the sample spectral function. In all calculations, convergence is ensured by increasing the resolution of both grids.

For the pairing potentials, we adopt the following pairing function
\begin{equation}
    \Delta_{p_y}=\Delta_0\left[\cos(\sqrt{3} \tilde{k}_x/2)\sin(\tilde{k}_y/2)+\sin(\tilde{k}_y)\right] \quad \Delta_{p_x}=\Delta_0\sin(\sqrt{3}\tilde{k}_x/2)\cos(\tilde{k}_y/2),
    \label{eq:pairings}
\end{equation}
with $\tilde{k}_{x/y}=2\pi \frac{k_{x/y}}{|k_{M,y}|}$ and $k_{M,y}$ the size of the y-component of the reciprocal moiré vector~\cite{schafferpairing}. These functions give $p_{y}$ and $p_x$ symmetry around the $\gamma$ point, and are periodic on the mBZ.
\section{Detecting nodal point using the tip Fermi Surface}
\label{app:nodal_detect}
\begin{figure}[h]
  \centering
\includegraphics[width=0.3\columnwidth]{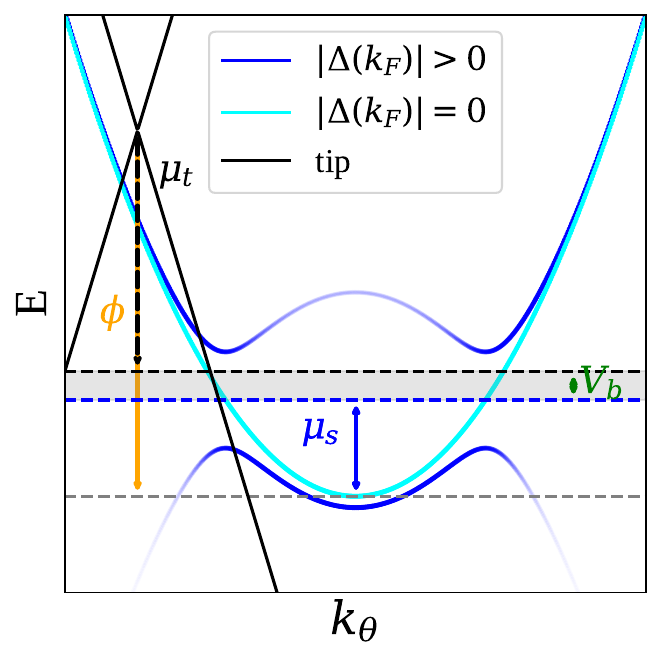}
  \caption{Schematic diagram of parabolic band structure with finite (blue) and zero (cyan) superconducting pairing. The color intensity denotes the spectral weight. The small bias voltage ($V_b$) opens the small gray window, in which tunneling is allowed. The tip chemical potential ($\mu_T$) and the electrostatic shift ($\phi$) are varied, while maintaining $V_b\approx0$. For the illustrated $\mu_T$, the tip bands cross only the cyan bands at the allowed window for tunneling (gray area). Thus, the only contribution to the current originates from this intersection.}
  \label{fig:nodalband}
\end{figure}

In this section, we show how additional information about the pairing magnitude can be extracted across the entire momentum space. This is achieved by exploiting the tunability of the tip chemical potential $\mu_T$, which allows for the controlled expansion of its Fermi circle. We focus on measurements at $V_b \approx 0$, where the tip Fermi level lies near the center of the superconducting gap at $\mu_S$ (see~\cref{fig:nodalband}). Under this condition, $\phi$ is determined by $\mu_T$ and $\mu_S$. At low temperatures, fixing $V_b \approx 0$ ensures that tunneling occurs only between states on the tip Fermi circle and states near the middle of the superconducting gap (up to spectral-function and temperature broadening). Consequently, $\left.\frac{dI}{dV_b}\right|_{V_b=0}$ can be examined as a function of the Fermi-circle radius, controlled by $\mu_T$.

The first important distinction is between a fully gapped superconductor and one that contains nodes. 
A fully gapped superconductor has no excitations at the center of the gap, so the current at small bias is suppressed and originates only from the finite broadening of the spectral function and from states near the gap edge. 
In contrast, the presence of a nodal point within the gap allows a direct contribution to the current. 
Consequently, when the tip Fermi surface intersects a nodal point, a pronounced increase in the differential conductance is expected. 
As $\mu_T$ increases further and the intersection no longer exists, the conductance gradually decreases; thus, a peak in $\frac{dI}{dV_b}$ as a function of $\mu_T$ is expected.  

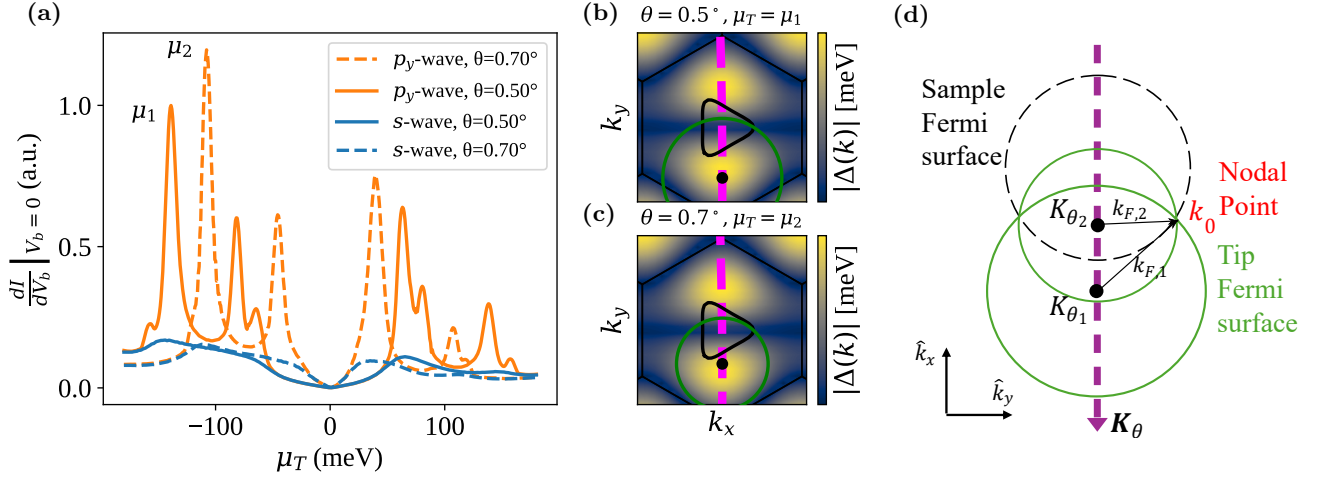
\begin{figure}[h]
    \centering
 \centering
  \subfloat{%
    \begin{tikzpicture}
      \node[inner sep=0] (img) {\includegraphics[width=0.65\columnwidth]{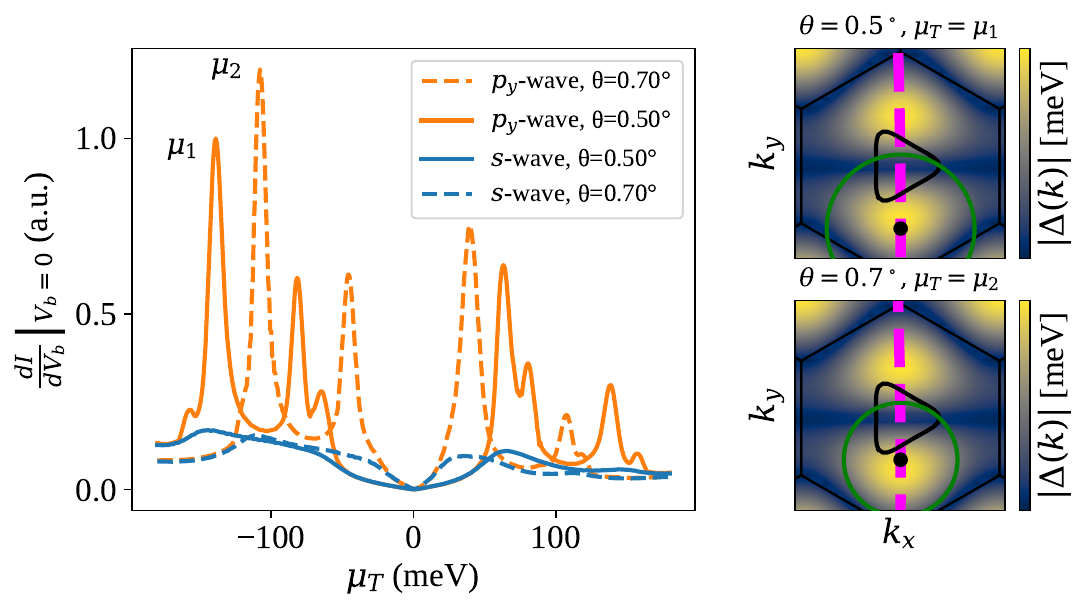}};
      \node[anchor=north west, xshift=8pt, yshift=0pt]  at (img.north west) {\textbf{(a)}};
      \node[anchor=north west, xshift=0.4\columnwidth+14pt, yshift=0pt]  at (img.north west) {\textbf{(b)}};
      \node[anchor=north west, xshift=0.4\columnwidth+14pt, yshift=-76pt]  at (img.north west) {\textbf{(c)}};
    \end{tikzpicture}%
    \label{fig:phiscan}%
  }%
\subfloat{\label{fig:nodal_inset1}}\hspace{0pt}%
\subfloat{\label{fig:nodal_inset2}}\hspace{0pt}%
  \subfloat{%
    \begin{tikzpicture}
      \node[inner sep=0] (img) {\includegraphics[width=0.35\columnwidth]{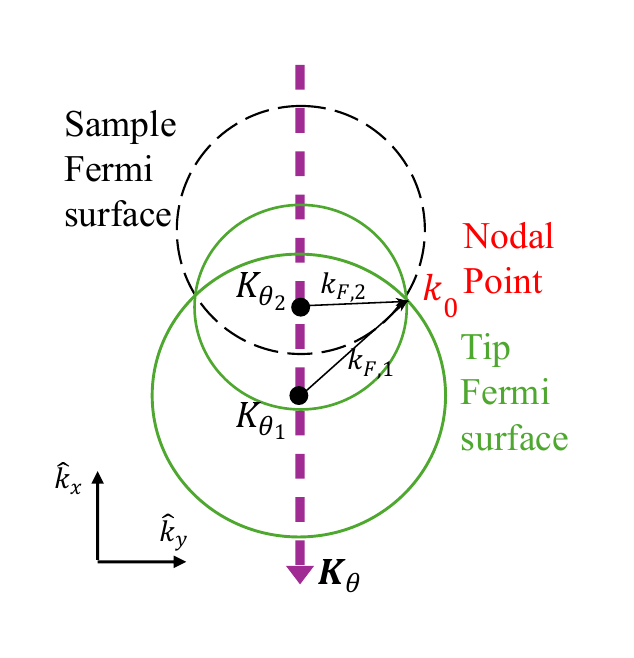}};
      \node[anchor=north west, xshift=4pt, yshift=0pt]  at (img.north west) {\textbf{(d)}};
    \end{tikzpicture}%
    \label{fig:nodalfermi}%
  }%
  \caption{(a) Calculated $dI/dV_{b}$ at $V_b=0$ as a function of $\mu_T$ for $s$-wave and $p_{y}$-wave pairings at two tip rotation angles. Solid lines correspond to $\theta=0.5^\circ$, dashed lines to $\theta=0.7^\circ$, with blue lines for $s$-wave and orange lines for $p_y$-wave pairing. The sharp features in the $p_y$ case arise from crossings between the tip Fermi circle and the nodal points of the sample. Two chemical potentials, $\mu_1$ and $\mu_2$, at which the zero-bias conductance is prominent, are indicated for the two rotation angles.  (b–c) Tip and sample Fermi surfaces (black and green contours) are shown for two sets of twist angles and tip chemical potentials, together with the magnitude of the $p_y$ pairing potential. A crossing occurs between the tip Fermi surface and a nodal point, where the pairing amplitude on the sample Fermi surface vanishes. The black dot marks the tip Dirac point location. The magenta dashed line indicates the tip line-scan. (d) Schematic illustration of nodal-momentum triangulation.  The black dashed line indicates the Fermi surface of the sample, with a nodal point at momentum $\mathbf{k}_0$. The Fermi circles of the tip at two rotation angles, $\theta_1$ and $\theta_2$, are shown in green. The radial momentum distances $k_{F,1}$ and $k_{F,2}$ between the nodal point and the tip Dirac points at $\mathbf{K}_{\theta_1}$ and $\mathbf{K}_{\theta_2}$ are marked by black arrows. The intersection of the two tip Fermi circles allows one to determine the nodal momentum geometrically.
}
\label{fig:nodal_loc}
\end{figure}

We use the same numerical framework as in the main text to demonstrate this effect, as shown in \cref{fig:phiscan}. The calculations are based on the BM model for MATBG, comparing $s$-wave and nodal $p_y$-wave pairings. As in the main text, the system is electron-doped so that the Fermi surface forms small pockets around the $\gamma$ point. As seen in \cref{fig:phiscan}, the $p_y$ pairing exhibits sharp features in $\frac{dI}{dV_b}$, whereas the $s$-wave pairing does not. We select two representative sharp features in the conductance for the nodal pairing and show the corresponding intersections between the tip and the sample Fermi surfaces that give rise to these features in \cref{fig:nodal_inset1,fig:nodal_inset2}.

Because the Fermi surface of the tip is well defined as a circle around the tip Dirac point, the momentum distance from the Dirac point can be directly inferred from  $k_F \equiv \frac{|\mu_T|}{\hbar v_D}$.
Repeating the zero-bias conductance measurement for two different tip rotation angles, $\theta_1$ and $\theta_2$, where the tip Dirac points are centered at $\mathbf{K}_{\theta_1}$ and $\mathbf{K}_{\theta_2}$, respectively, allows one to geometrically triangulate the nodal-point momentum via  
\begin{align}
k_{0,x} &= 
\frac{k_{F,1}^{2}-k_{F,2}^{2}+(K_{\theta_2,x}-K_{\theta_1,x})^{2}}
{2\,(K_{\theta_2,x}-K_{\theta_1,x})}, \\
k_{0,y} &= \pm \sqrt{\,k_{F,1}^{2}-\bigl(k_{0,x}-K_{\theta_1,x}\bigr)^{2}} ,
\end{align}
where $k_{F,1/2}$ denote the Fermi momenta, which are the distances between the tip Dirac points and the nodal point at $\theta_1$ and $\theta_2$. 
The two signs of $k_{0,y}$ correspond to symmetric solutions mirrored across the line connecting $\mathbf{K}_{\theta_1}$ and $\mathbf{K}_{\theta_2}$. 
The schematic geometry of this procedure is shown in \cref{fig:nodalfermi}.  This triangulation method is independent of the sample’s Fermi-surface shape, band structure, and specific pairing symmetry.
Importantly, because the QTM simultaneously probes three $C_{3z}$-related tunneling trajectories, each solution is indistinguishable under $120^{\circ}$ rotations. 
Altogether, the nodal momentum can therefore be located at up to six symmetry-related points.
In addition, in \cref{fig:phiscan} the positions of the peaks are symmetric under the transformation $V_b \rightarrow -V_b$, 
which corresponds to tunneling into the valence and conduction bands of the MLG tip. However, peak intensities depend on the sign of $V_b$, due to the different overlaps and tunneling matrix elements 
between the tip and sample bands in the two cases.

\clearpage
\makeatletter
\begin{@fileswfalse}
\bibliography{refs}
\end{@fileswfalse}
\makeatother

\end{document}